\documentclass[aps,prb,reprint,groupedaddress,showpacs]{revtex4-1}

\pdfoutput=1

\usepackage{amsmath}
\usepackage{amsfonts}
\usepackage{amssymb}
\usepackage{graphicx}
\usepackage[caption=false]{subfig}

\newcommand{\ket}[1]{|#1\rangle}

\newcommand{\beq}{\begin{equation}}  
\newcommand{\eeq}{\end{equation}}  

\begin{document}


\title{Landau-Zener dynamics of a nanoresonator containing a tunneling spin}


\author{Michael F. O'Keeffe, Eugene M. Chudnovsky, and  Dmitry A. Garanin}
\affiliation{Physics Department, Lehman College, City University of New York,
    250 Bedford Park Boulevard West, Bronx, New York, 10468-1589, USA}


\date{\today}

%
%
\begin{abstract}
We study the Landau-Zener dynamics of a tunneling spin coupled to
a torsional resonator.  For strong spin-phonon coupling, when the
oscillator frequency is large compared to the tunnel splitting,
the system exhibits multiple Landau-Zener transitions.
Entanglement of spin and mechanical angular momentum results in
abrupt changes of oscillator dynamics which coincide in time with
spin transitions. We show that a large number of spins on a single
oscillator coupled only through the in-phase phonon field behaves
as a single large spin, greatly enhancing the spin-phonon
coupling. We compare purely quantum and semiclassical dynamics of
the system and discuss their experimental realizations. An experiment
is proposed in which the field sweep is used to read out the exact
quantum state of the mechanical resonator.
\end{abstract}

\pacs{75.80.+q, 75.45.+j, 75.50.Xx, 85.65.+h}

\maketitle


%
%
\section{Introduction \label{sec:intro}}

The Landau-Zener model \cite{Landau:1932, *Zener:1932,
*Stueckelberg:1932, *Majorana:1932} describes a two-state system
in which the bias between diagonal states varies linearly with
time as they are swept through an avoided crossing.  It is one of
the few practically important time-dependent Hamiltonians for
which the Scrh\"odinger equation is exactly solvable.
The Landau-Zener method has recently found a natural application
in the experimental characterization of single molecule magnets
\cite{Wernsdorfer:1999}.  Theoretical studies of Landau-Zener
transitions in nanomagnets have included many-body effects
\cite{Garanin:2005} and superradiance \cite{Chudnovsky:2002, *Chudnovsky:2004}.
Some important theorems have been proven about generalizations of
the Landau-Zener problem\cite{Brundobler:1993, *Sinitsyn:2004},
and certain multilevel cases have been exactly solved
\cite{Sinitsyn:2012}.  It has been used as a model for the
dynamics of quantum phase transitions \cite{Zurek:2005}, and
topological defect formation \cite{Damski:2005}.

A natural extension of the two-level quantum physics is the
two-level system coupled to one or several quantized modes of a
harmonic oscillator. Studies of Landau-Zener oscillator dynamics
have probed coherent \cite{Keeling:2008, Sun:2012}, dissipative
\cite{Wubs:2006, Saito:2007}, and temperature-dependent
\cite{Whitney:2011} effects.  Landau-Zener interferometry has
provided a quantitative measure of coupled dynamics
\cite{Shevchenko:2010} and has been experimentally verified in the
nanomechanical measurement of a superconducting qubit
\cite{LaHaye:2009}.

The Landau-Zener effect in spin systems should be considered in
conjunction with the transfer of angular momentum manifested in
the Einstein - de Haas effect. This effect has allowed precision
measurement of the magneto-mechanical ratio of a thin
ferromagnetic film on a microcantilever \cite{Wallis:2006}.
Torsional oscillators have been used as precision torque
magnetometers in nanomechanical detection of itinerant electron
spin-flip at a ferromagnet-normal metal junction
\cite{Zolfagharkhani:2008} and measurement of phase transitions of
small magnetic disks in and out of the vortex state
\cite{Davis:2010}.  Semiclassical models of Landau-Zener dynamics
have been developed to describe magnetic molecules coupled to
mechanical resonators and bridged between conducting leads
\cite{Jaafar:2010, Jaafar:2011}. A full quantum treatment of the
interaction between a single spin and a torsional oscillator has
recently been developed \cite{Kovalev:2011, Garanin:2011}.

Realizing a quantum magneto-mechanical system with strong
spin-phonon coupling has been an experimental challenge.  A recent
experiment \cite{Ganzhorn:2013} has shown the first evidence of
strong spin-phonon coupling in a single molecule magnet grafted
onto a carbon nanotube.  Spin reversal of the single molecule
magnet during a Landau-Zener sweep coincides with an abrupt
increase in the differential conductance through the carbon
nanotube.  This has been interpreted as the spin transition
exciting a longitudinal stretching mode of the carbon nanotube,
which enhances electron tunneling from the lead onto the nanotube
through electron-phonon coupling.

We propose multiple schemes to realize strongly coupled dynamics
of a tunneling macrospin with torsional oscillations of a
nanoresonator in a Landau-Zener experiment.  We investigate the
Landau-Zener dynamics of a tunneling spin coupled to a torsional
oscillator, using a fully quantum mechanical model.  The
oscillator could be a torsional paddle resonator, a
microcantilever, a carbon nanotube, or a single magnetic molecule
between two point contacts. The tunneling spin could be a single
molecule magnet, an ensemble of single molecule magnets, or a
single-domain ferromagnetic particle with strong uniaxial
anisotropy. For a collection of single molecule magnets placed on
a torsional resonator or cantilever far apart from each other that
they are not directly coupled through dipole interactions, we
develop a semiclassical model of magnetization dynamics.  We
predict superradiant enhancement \cite{Chudnovsky:2004} of the spin-phonon coupling
for this ensemble system.  Comparison of these two models shows
their correspondence.

The coupling between spin and mechanical angular momentum is
mandated by the conservation of total angular momentum
$\bf J = S + L$, with $\bf L$ beingthe mechanical angular momentum.
In a free particle, when a spin
tunnels from $\bf S$ to $\bf - S$, the particle must change its
mechanical angular momentum $\bf L$. This changes its kinetic
energy by an amount of order $\hbar^2 \mathbf{S}^2 / I$, where
$I_z$ is the moment of inertia about the rotation axis. For a macroscopically large body,
the large moment of inertia makes this rotational kinetic energy
negligibly small. But for a small particle this can become
comparable to the energy gain $\Delta$ due to tunnel splitting
\cite{Chudnovsky:1994}. The ratio of these two quantities, the
magneto-mechanical ratio $ \alpha = 2 \hbar^2 S^2 / I_z \Delta$
determines the ground state of the system for a free particle
\cite{Chudnovsky:2010, *OKeeffe:2012}.  For large particles
$\alpha \ll 1$ and the ground state is the well-known tunnel split
state $\Psi \sim \ket{\psi_{S}} + \ket{\psi_{-S}}$. For small
particles, such as the Fe$_8$ single molecule magnet with $I_z \sim
10^{-42}$ kg$\cdot$m$^2$, $\alpha \gg 1$ and spin tunneling is
suppressed as the spin localizes in either direction along the
easy axis.

Similar effects arise in systems that undergo torsional
oscillations. Examples are a single molecule magnet bridged
between conducting leads, a nanomagnet attached to a carbon
nanotube bridge, or a nanomagnet coupled to a resonator such as a
torsional paddle oscillator or microcantilever.  The mechanical
resonance occurs at a frequency $\omega_r = \sqrt{k/I_z}$, where $k$
is the effective stiffness against the linear restoring torque and
$I_z$ is the moment of inertia of the nanomagnet-resonator
combination. A convenient measure of the effect of oscillations is
the dimensionless parameter $r = \hbar \omega_r / \Delta$, the
ratio between the oscillator energy to tunnel splitting.  As we
will see in Sec. \ref{sec:model}, the coupling between
magnetization and oscillator dynamics is given by the factor
$\lambda = \sqrt{\alpha / r} = \sqrt{2 \hbar S^2 / I_z \omega_r}$. The
most interesting effects occur for strong coupling $\lambda \sim
1$ and oscillator frequency much larger than tunnel splitting $r
\gg 1$. A large spin, small moment of inertia, and weak torsional
spring constant are required for strong coupling.
\begin{figure}[h!]  
    \subfloat{    \label{fig:CNT_SMM}
        \includegraphics[trim = 1.5cm 4cm 1cm 4cm, clip, width=0.24\textwidth]{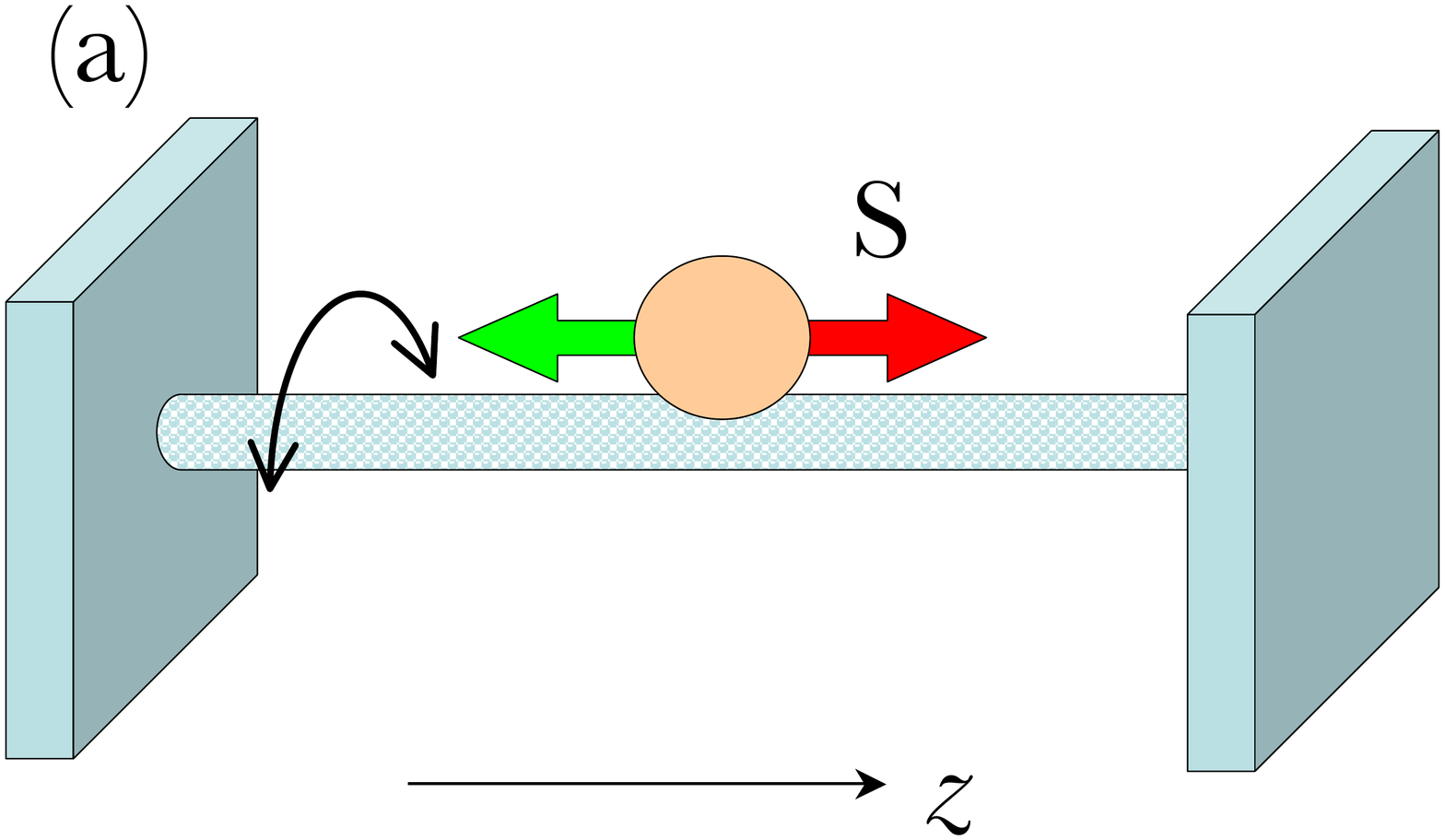}}
    \subfloat{        \label{fig:cantilever_SMMs}
        \includegraphics[trim = 1cm 4cm 1cm 4cm, clip, width=0.24\textwidth]{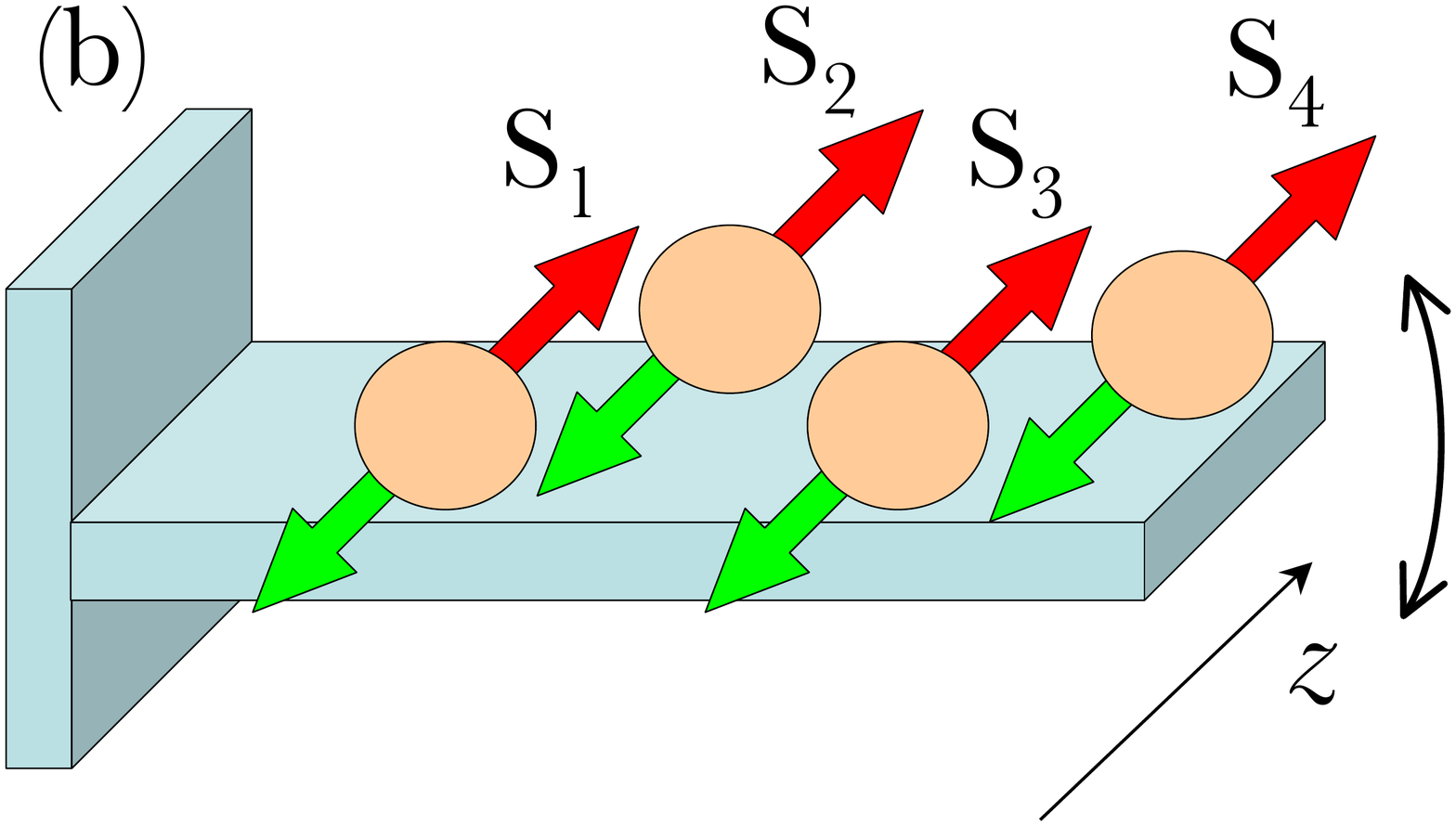}}
    \caption{Possible experimental geometries described by the models studied in
    this paper. In both cases the easy axis of the macrospin coincides with the
rotation axis of the oscillator. (a) Single molecule magnet
grafted on a carbon nanotube.  (b) Ensemble of single molecule
magnets on a nanocantilever. }
    \label{fig:spin_osc}
\end{figure}
 article is organized as follows.  In Sec. \ref{sec:model} we
briefly review the Landau-Zener model, and construct the quantum
mechanical model of a spin coupled to a torsional resonator with
an external magnetic field that varies linearly in time.  Sec.
\ref{sec:LZSO_dynamics} contains numerical and analytical results
of the fully quantum spin dynamics for a variety of parameter
ranges. Oscillator dynamics are presented in Sec. \ref{sec:osc}. A
semiclassical model of superradiant dynamics in an ensemble of
spins on a single resonator is developed in Sec. \ref{sec:SR}.
Finally, we discuss the interpretation of our results for various
experimental realizations in Sec. \ref{sec:conclusion}.

%
%
\section{Model \label{sec:model}}

%
\subsection{Landau-Zener Transitions in a Two-State System\label{subsec:LZ}}

\begin{figure}[h]   
    \centering
    \subfloat{ \label{fig:E_LZ} \includegraphics[trim = 0cm 0cm 0cm 0cm, clip, width=0.23\textwidth]{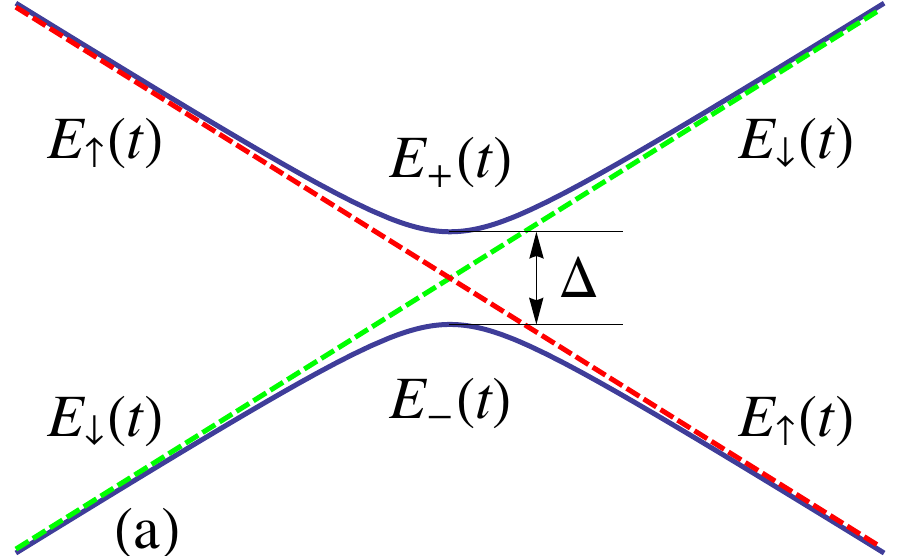}}
    \subfloat{\label{fig:P_LZ} \includegraphics[trim = 0cm 0cm 0cm 0cm, clip, width=0.23\textwidth]{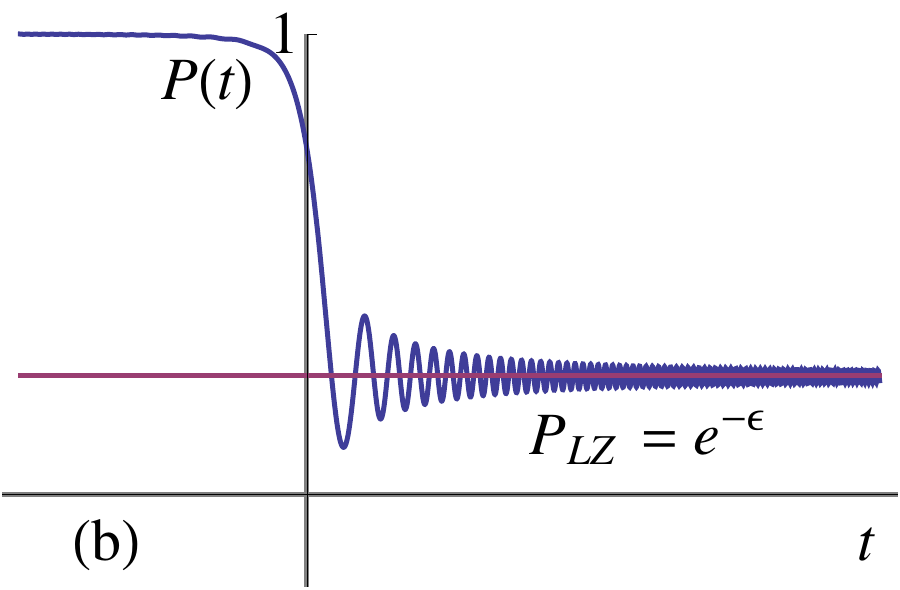}}
    \caption{(a) Adiabatic $E_\pm (t)$ and diabatic $E_{\uparrow \downarrow} (t)$ energy levels of the LZ Hamiltonian as a function of time. (b) Probability $P(t)$ of staying in the initial $\ket \downarrow$ state as a function of time, and asymptotic staying probability $P_{LZ}$.  }
    \label{fig:LZ}
\end{figure}

We review relevant features of the Landau-Zener model, which describes a two-level system driven by a classical field that varies linearly in time.  The LZ  Hamiltonian is
\beq        \label{eq:H_LZ}
    \hat H_{LZ} = - \frac{vt}{2} \sigma_z - \frac \Delta 2 \sigma_x,
\eeq
in terms of Pauli matrices $\sigma_z$ and $\sigma_x = \sigma_+ + \sigma_-$, where $v$ is the sweep rate and $\Delta$ is the tunnel splitting.  Diabatic states $\ket \uparrow$ and $\ket \downarrow$ are eigenstates of $\sigma_z$ with diabatic energies $E_{\uparrow \downarrow} (t) = \pm vt / 2$, which are the linear functions in Fig.~\ref{fig:E_LZ}.  We take the sweep rate $v$ positive, so the positive (negative) sign corresponds to spin down (up).  For nonzero $\Delta$, the diabatic states are not eigenstates of the Hamiltonian.  Diagonalizing $\hat H_{LZ}$ gives adiabatic energies
\beq
    E_\pm(t) = \pm \frac 1 2  \sqrt{ (v t)^2 + \Delta^2}
\eeq
which are the upper and lower curves in Fig.~\ref{fig:E_LZ} with splitting $\Delta$ at $t = 0$.  The corresponding adiabatic eigenstates $\ket{+}$ and $\ket{-}$ are
\beq
    \ket \pm = \frac{1}{\sqrt 2} (C_\mp \ket \uparrow \mp C_{\pm} \ket \downarrow),
\eeq
where $C_\pm$ depend explicitly on time,
\beq
    C_\pm =  \sqrt{1 \pm \frac{vt}{\sqrt{(vt)^2 + \Delta^2} } }.
\eeq
For times $|t| \gg \Delta / v$ the adiabatic states asymptotically coincide with the diabatic states.

The state of the system
\beq
    \Psi(t) = c_\uparrow (t) \ket \uparrow + c_\downarrow (t) \ket \downarrow
\eeq
evolves according to the time-dependent Schr\"odinger equation
\beq
    i \hbar \frac{\partial \Psi}{\partial t} = \hat H \Psi
\eeq
with initial conditions $c_\uparrow (-\infty) = 0$, $|c_\downarrow (-\infty)| = 1$.
After eliminating $c_\downarrow$, we obtain the second order differential equation
\beq
    \ddot c_\uparrow (t) + \left [ \left ( \frac{\Delta}{2 \hbar} \right )^2 - \frac{i v}{2 \hbar}
        + \left ( \frac{v t}{2 \hbar} \right ) \right ] c_\uparrow (t) = 0
\eeq
which can be put into the standard form of the Weber equation.
The exact solution \cite{Zener:1932} gives
\beq
    c_\uparrow (t) = \sqrt \gamma e^{- \pi \gamma / 4} D_{-\nu-1} (-iz)
\eeq
where
\beq
    \gamma = \frac{\Delta^2}{4 \hbar v}, \qquad
    \nu = i \gamma, \qquad
    z = \sqrt{\frac{v}{\hbar}} e^{-i \pi / 4} t,
\eeq
and $D_{-\nu-1} (-iz)$ are parabolic cylinder functions.  The staying probability for the spin-down state as function of time is $P(t) = |c_\downarrow (t)|^2$.  The exact asymptotic limit for $t = \infty$, known as the Landau-Zener probability, is
\beq        \label{eq:P_LZ}
    P_{LZ} = e^{- \epsilon}, \qquad  \epsilon = \frac{\pi \Delta^2}{2 \hbar v}.
\eeq
$P(t)$ and $P_{LZ}$ are shown in Fig.~\ref{fig:P_LZ}.  The same $P(t)$ and $P_{LZ}$ can be obtained from the Heisenberg equations of motion for $\langle \sigma_z (t) \rangle$.
An intuitive understanding of the Landau-Zener transition comes from considering the time spent in the tunneling region between adiabatic states and the tunneling time between these states.  Let $\tau_{LZ} \sim \text{max}(\sqrt{\hbar / v}, \Delta / v)$ be the time spent in the tunneling region and $\tau_\Delta \sim \hbar / \Delta$ be the tunneling time at the crossing.  The Landau-Zener exponent is proportional to the ratio of these times $\epsilon \sim \tau_{LZ} / \tau_\Delta$.  For a slow sweep the system will evolve adiabatically, spending long enough in the tunneling region that it will continually relax to the ground state, making $\epsilon \gg 1$ and $P_{LZ} \rightarrow 0$.  In the opposite limit, a fast sweep through the tunneling region makes $\epsilon \ll 1$ and the staying probability saturates at $P_{LZ} \rightarrow 1$.

%
\subsection{Landau-Zener Transitions in a Spin-Oscillator System\label{subsec:LZSO}}

Consider a tunneling spin which is projected onto the lowest tunneling doublet.  This spin is coupled to a torsional nanoresonator with rigidity $k$ that can rotate about the $z$-axis, which coincides with the easy axis of the spin.  The Hamiltonian is \cite{Kovalev:2011, Garanin:2011},
\beq        \label{eq:H_SO}
    \hat H = \frac{\hbar^2 L_z^2}{2 I} + \frac{I_z \omega_r^2 \phi^2}{2}
        - \frac{W (t)}{2} \sigma_z - \frac {\Delta}{2} \left (e^{- i 2 S \phi} \sigma_+ + e^{ i 2 S \phi} \sigma_- \right ).
\eeq
The fundamental frequency of torsional oscillations is $\omega_r = \sqrt{k/I_z}$, where $I_z$ is the moment of inertia of the resonator about its rotation axis.  An external longitudinal magnetic field $B_z (t)$ applied along this axis creates a time-dependent energy bias $W(t) = 2 S g \mu_B B_z (t)$. The Landau-Zener problem describes a linear field sweep, $W(t) = v t$.  The operator of mechanical angular momentum, $L_z = - i \partial_\phi$, and the angular displacement $\phi$ of the oscillator obey the usual commutation relation $[\phi, L_z] = i$.

The last term in the Hamiltonian describes the entanglement
between spin transitions and mechanical rotations.  A typical
single molecule magnet has a large spin and strong uniaxial
anisotropy, producing a zero-field splitting between degenerate
ground states $\ket{\psi_{\pm S}}$ pointing in either direction
along the easy axis.  Any symmetry breaking interactions, such as
transverse anisotropy or an external field, break this degeneracy
producing tunnel split states $\Psi \sim \ket{\psi_{S}} \pm
\ket{\psi_{-S}}$ which are represented by the pseudospin
$\boldsymbol \sigma$.  The tunnel splitting $\Delta$ is generally
many orders of magnitude less than the energy to the next spin
level.  In the case of the spin-10 single molecule magnet Fe$_8$,
the crystal field Hamiltonian describing the magnetic anisotropy
is $\hat H_S = - D \hat S_z^2 + d \hat S_y^2$, with $d \ll D$.
Full perturbation theory \cite{Garanin:1991} gives \beq
    \Delta = \frac{8 S^{3/2}}{\pi^{1/2}} \left ( \frac{d}{4D} \right )^S D,
\eeq
where we can see that $\Delta \ll 2 S D$, which is the distance to the next spin level.  The crystal field Hamiltonian $\hat H_S$ is defined with respect to coordinate axes that are rigidly coupled to the molecule or crystal.  Because the particle is free to rotate, the crystal field Hamiltonian must be transformed to the fixed frame of the laboratory.  Projecting the crystal field Hamiltonian onto the lowest tunneling doublet, rotating to the lab frame using $\hat U(\hat S_z) = e^{i \hat S_z \phi}$, where $\hat S_z \ket{\psi_{\pm S}} \simeq \pm S \ket{\psi_{\pm S}}$, $\hat H'_S = \hat U \hat H_S \hat U^{-1}$ gives the final term of the Hamiltonian.

We now consider the spin-oscillator Hamiltonian with a linear field sweep $W(t) = v t$.  Introducing the usual annihilation and creation operators, $a$ and $a^\dagger$,
\beq
    \phi = \sqrt{\frac{\hbar}{2 I_z \omega_r}} (a^\dagger + a), \qquad
    L_z = i \sqrt{\frac{I_z \omega_r}{2 \hbar}} (a^\dagger - a)
\eeq
into Eq.~\eqref{eq:H_SO} gives
\beq        \label{eq:H_LZSO}
    \hat H = \hbar \omega_r a^\dagger a - \frac{vt}{2} \sigma_z - \frac \Delta 2 (e^{- i \lambda (a^\dagger + a)} \sigma_+ + e^{ i \lambda (a^\dagger + a)} \sigma_-),
\eeq
where we have dropped unessential constant terms.  We will find it useful to adopt dimensionless units $\hat H' = \hat H / \Delta$ and $t' = \Delta t / \hbar$,
\beq        \label{eq:H_LZSO_dimless}
    \hat H' = r a^\dagger a - \frac{v't'}{2} \sigma_z - \frac 1 2 (e^{- i \lambda (a^\dagger + a)} \sigma_+ + e^{ i \lambda (a^\dagger + a)} \sigma_-),
\eeq
which shows that the system depends on three parameters.  The parameters
\beq        \label{eq:lambda_r}
    \lambda = \sqrt{\frac{2 \hbar S^2}{I_z \omega_r}}, \qquad
    r = \frac{\hbar \omega_r}{\Delta}
\eeq
describe the spin-oscillator relationship.  $\lambda$ is the coupling strength between the spin and oscillator and $r$ is the ratio of mechanical oscillation to tunnel splitting frequency.  The relationship between $\lambda$ and $r$  can be understood by the so-called magneto-mechanical ratio,
\beq        \label{eq:alpha}
    \alpha = \lambda^2 r = \frac{2 \hbar^2 S^2}{I_z \Delta},
\eeq
which is the ratio of the change in rotational kinetic energy associated with a spin transition $\bf S \rightarrow - \bf S$ to the tunnel splitting energy.  The third parameter is the effective sweep rate $v'$, or equivalently the Landau-Zener exponent $\epsilon$ defined in Eq.~\eqref{eq:P_LZ},
\beq        \label{eq:v'}
    v' = \frac{\pi}{2 \epsilon} = \frac{\hbar v}{\Delta^2}.
\eeq

We choose the spin up/down basis for the two-level system and a Fock state basis for the harmonic oscillator.  A direct product of these two bases will form the basis of the spin-oscillator system.  Matrix elements of the Hamiltonian Eq.~\eqref{eq:H_LZSO} are
\begin{align}
    H_{m \sigma, n \sigma'} &= \left ( \hbar \omega_r m - \frac{v t}{2} \sigma \right ) \delta_{mn} \delta_{\sigma \sigma'}  \\
        & \qquad - \left [ \frac{\Delta_{mn}}{2} \delta_{\sigma, -1} \delta_{\sigma',1} + \frac{\Delta^*_{mn}}{2} \delta_{\sigma, 1} \delta_{\sigma', -1} \right ], \nonumber
\end{align}
where $\sigma = -1, 1$ corresponds to spin down and up states, respectively.  The full Fock space has an infinite number of states, although we will use a truncated basis for numerical computations.  Tunneling matrix elements
\beq        \label{eq:Delta_mn}
    \Delta_{mn} = \Delta \, \kappa_{mn} (\lambda),
\eeq
depend on the coupling $\lambda$ through matrix elements of the displacement operator $\hat D (\xi) = \exp (\xi a^\dagger - \xi^* a)$, $\xi = -i \lambda$,
\beq        \label{eq:kappa_mn}
    \kappa_{mn} (\lambda) = e^{-\lambda^2 / 2} (-i \lambda)^{m-n} \sqrt{\frac{n!}{m!}} L_n^{(m-n)} (\lambda^2)
\eeq
for $ m \geq n$, and $m \leftrightharpoons n$ for $m < n$.  $L_n^{(m-n)} (x)$ are generalized Laguerre polynomials, and the real parameter $\lambda$ is defined in Eq.~\eqref{eq:lambda_r}.  The first few $\kappa_{mn}$ are
\begin{align}
    &\kappa_{00} = e^{-\lambda^2 / 2}, \qquad \kappa_{01} = \kappa_{10} = - i \lambda e^{-\lambda^2 / 2}, \nonumber  \\
    &\kappa_{11} = (1 - \lambda^2) e^{-\lambda^2 / 2}.
\end{align}

%
%
\section{Landau-Zener Spin-Oscillator Dynamics \label{sec:LZSO_dynamics}}

%
\subsection{Adiabatic energy levels \label{subsec:adiabatic}}
\begin{figure}[h]   
    \includegraphics[trim = 0cm 0cm 0cm 0cm, clip, width=0.47\textwidth]{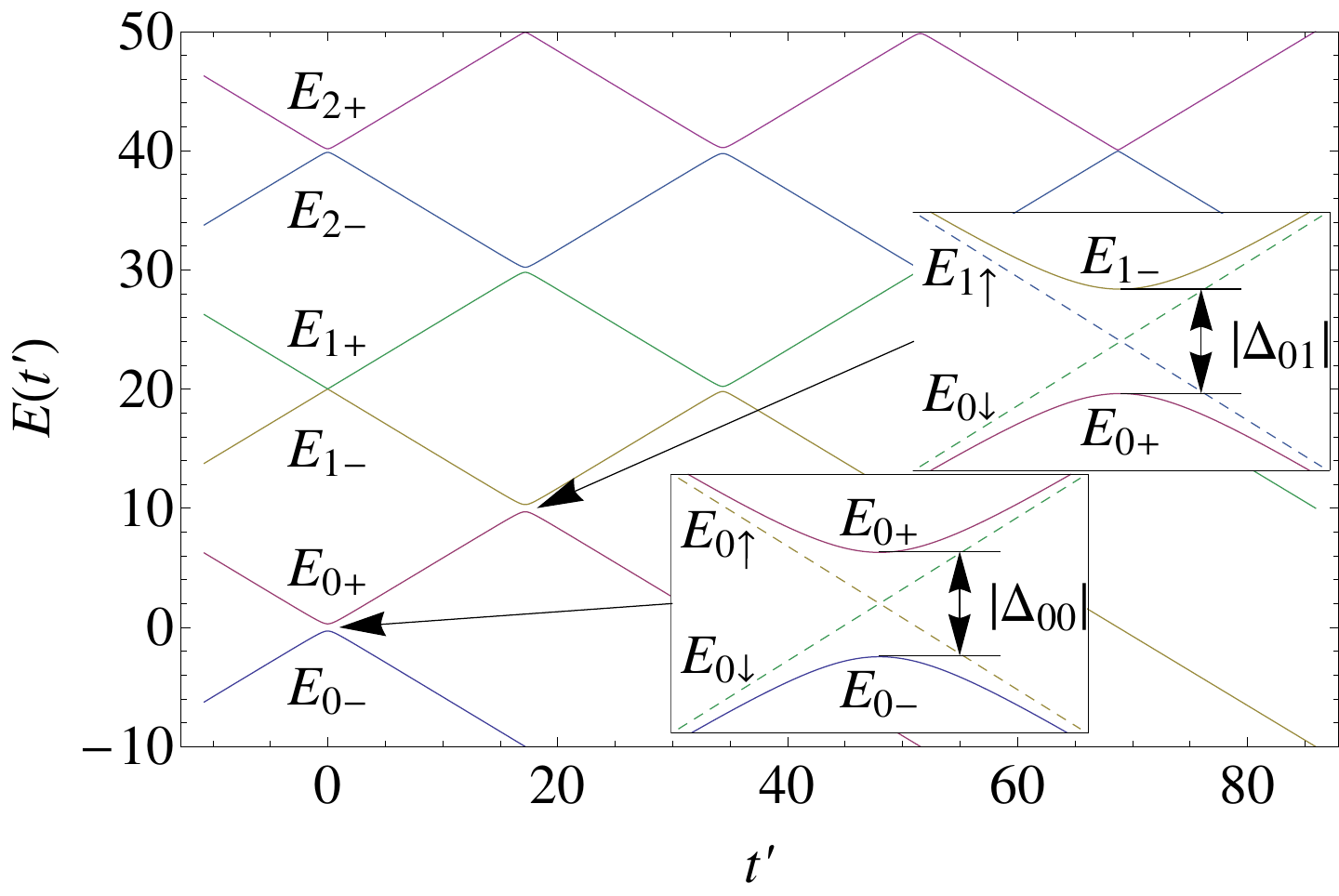}
    \caption{Energy (in units of $\Delta$) as a function of time for $r=20$, $\lambda=1$, $\epsilon=1.35$.  Solid lines are adiabatic energy levels $E_{n \pm}$, and diabatic energies $E_{n \downarrow \uparrow}$ are dashed lines in the insets.  Crossings occur at $t_k$.}
    \label{fig:adiabatic}
\end{figure}

Numerically solving $\det(\hat H - E I) = 0$ gives the adiabatic energy levels  $E_{n \pm}$, shown in Fig.~\ref{fig:adiabatic}.  Diabatic energy levels $E_{n \downarrow \uparrow}$, dotted lines in the insets of Fig.~\ref{fig:adiabatic},  are eigenvalues of the noninteracting part of the Hamiltonian (the first two terms in Eq.~\eqref{eq:H_LZSO}), given by
\beq        \label{eq:E_n_updown}
    \frac{E_{n \downarrow \uparrow}}{\Delta} = n r \pm \frac{v' t'}{2}.
\eeq
The spin down (up) states have positive (negative) slopes with $y$-intercepts $n \omega$. Diabatic energies $E_{n \downarrow}$ and $E_{m \uparrow}$ cross at times
\beq        \label{eq:t_k}
    t'_k = k \frac{r}{v'}, \qquad k = m - n \in \mathbb Z.
\eeq
When the oscillator frequency is much larger than the sweep rate, $r \gg v'$, the transitions are independent.  Note that the indices on the adiabatic and diabatic energies only coincide near $t = 0$, but will in general be different after successive crossings.  The tunnel splittings $|\Delta_{mn}|$ between adiabatic states occur at the crossing of diabatic energies $E_{m \downarrow}$ and $E_{n \uparrow}$, and depend on the coupling strength through Eqs. \eqref{eq:Delta_mn} and \eqref{eq:kappa_mn}.  When $r \gtrsim v'$, successive transitions occur within short times of each other.  Once $r \lesssim v'$ there are many closely spaced levels near $t = 0$.

Consider a single spin initially spin-down with the oscillator in the zero phonon state, i.e. $\Psi(t = - \infty) = \ket 0 \ket \downarrow$.  The system is initially in the adiabatic energy state $E_{0-}$ which corresponds to the diabatic state $E_{0 \downarrow}$.  At $t_0 = 0$, diabatic states $E_{0 \downarrow}$ and $E_{0 \uparrow}$ cross, and adiabatic states $E_{0-}$ and $E_{0+}$ approach each other with minimum separation $|\Delta_{00}| = \Delta e^{-\lambda^2 / 2}$.  If the spin remains in the initial adiabatic state $E_{0-}$ after the avoided crossing, it flips and will see no more possible transitions, as $E_{0-}$ coincides with $E_{0 \uparrow}$ long after the avoided crossing at $t_0$.  If the spin does not flip, it will follow the adiabatic state $E_{0+}$ which coincides with $E_{0 \downarrow}$ long after $t_0$.  The next crossing between diabatic states $E_{0 \downarrow}$ and $E_{1 \uparrow}$ occurs at $t_1$, with tunnel splitting $|\Delta_{01}| = \Delta e^{- \lambda^2 / 2} \lambda$ between diabatic states $E_{0+}$ and $E_{1-}$.  Remaining in the adiabatic state $E_{0+}$ will coincide with $E_{1 \uparrow}$ for times long after $t_1$.  If the spin does not flip at $t_1$, the system will remain in the $E_{1-}$ adiabatic state, coinciding with $E_{0 \downarrow}$ long after $t_1$.  In general the crossing between state $\ket 0 \ket \downarrow$ and $\ket k \ket \uparrow$ occurs at $t_k$ with splitting $|\Delta_{0k}| = \Delta e^{-\lambda^2 / 2} | \kappa_{0k} (\lambda) |$.  Notice that the avoided crossing between $E_{1-}$ and $E_{1+}$ at $t_0 = 0$, given by $|\Delta_{11}| = \Delta e^{-\lambda^2/2} | 1 - \lambda^2 |$ does exactly go to zero when $\lambda = 1$.

%
\subsection{Strong coupling \label{subsec:strong}}
\begin{figure}[h!]    
        \subfloat{\label{fig:P_r20_lambda1_ground}\includegraphics[trim = 0cm 0cm 0cm 0cm, clip, width=0.45\textwidth]{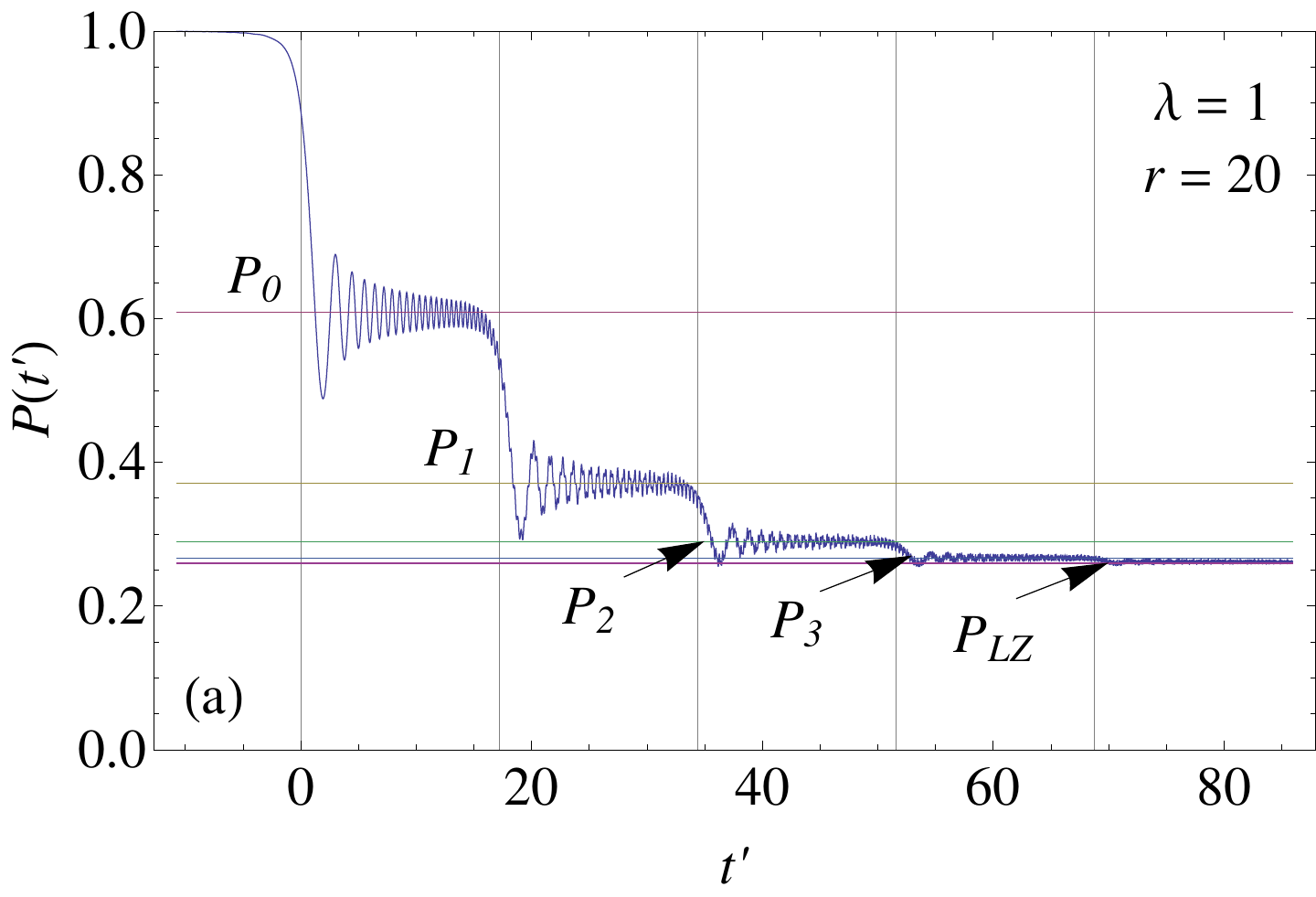}}  \\
        \subfloat{\label{fig:P_r20_lambda2_ground}\includegraphics[trim = 0cm 0cm 0cm 0cm, clip, width=0.45\textwidth]{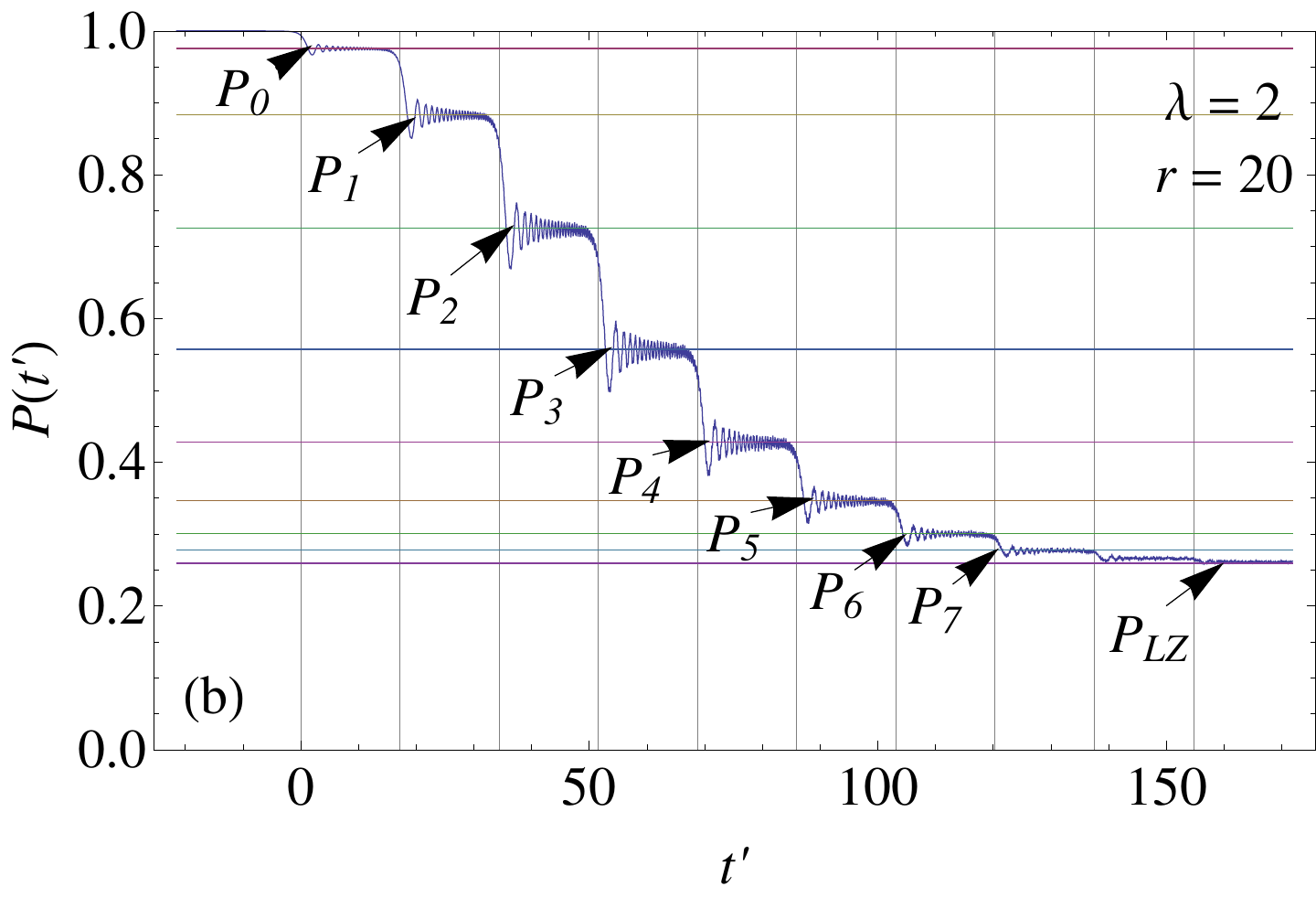}}  \\
        \subfloat{\label{fig:P_r1_lambda1_ground}\includegraphics[trim = 0cm 0cm 0cm 0cm, clip, width=0.45\textwidth]{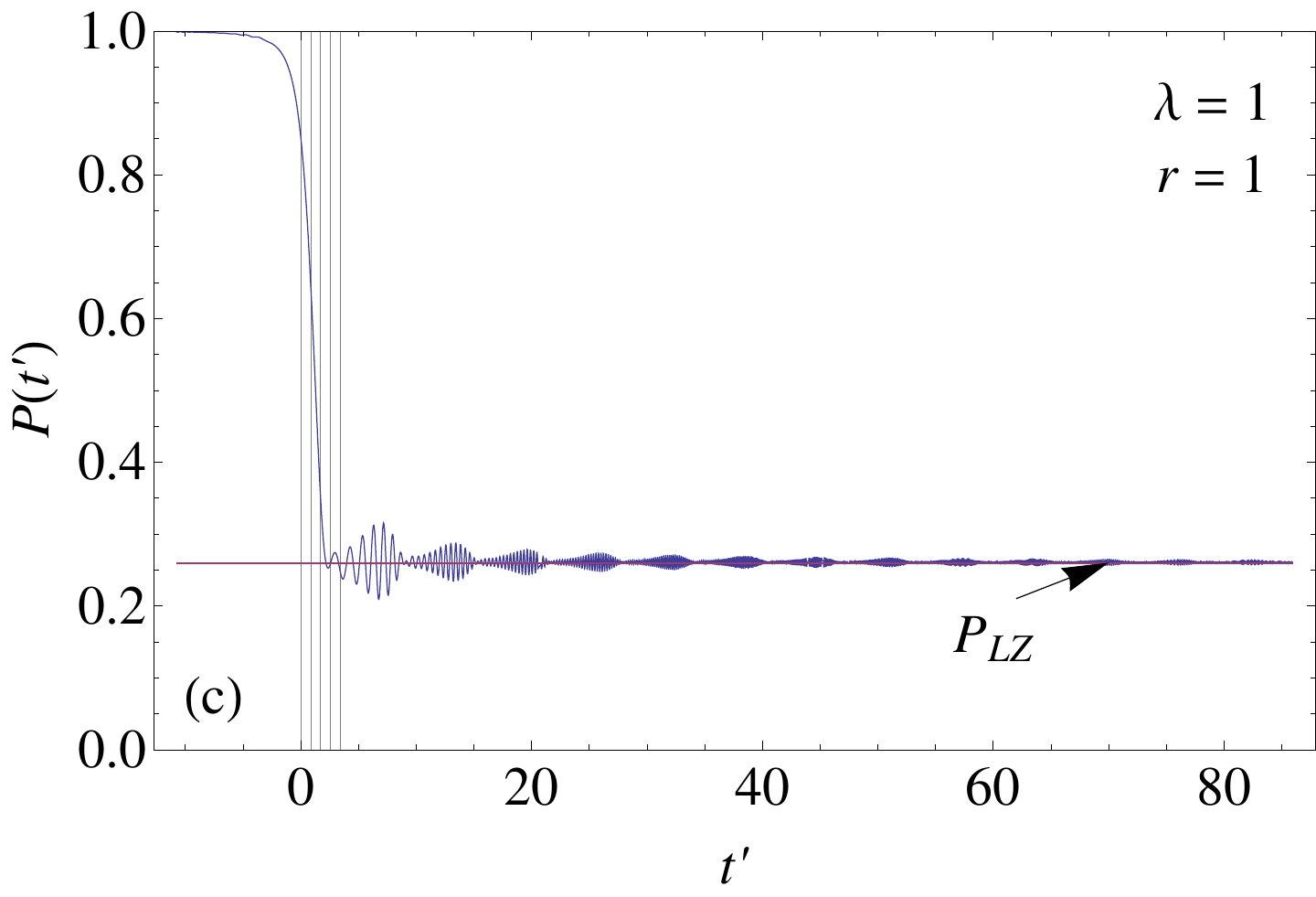}}
        \caption{Time dependence of the probability that the spin stays in the initial spin-down state for initial state $\Psi(- \infty) = \ket 0 \ket \downarrow$ with $\epsilon = 1.35$.  Vertical lines at $t_k$ denote avoided crossing of adiabatic energy levels.  Horizontal lines are exact results $P_N$ for independent transitions.}
    \label{fig:P_ground}
\end{figure}
We study the dynamics of the spin-oscillator system for various parameter ranges.  Expanding the wave function of the system in this basis
\beq
    \ket{\Psi (t)} = \sum_{m=0}^\infty \sum_{\sigma = \pm 1} C_{m \sigma} (t) \ket m \ket \sigma,
\eeq
the time-dependent Schr\"odinger equation yields the system of coupled differential equations,
\begin{align}
    i \frac{d C_{m, \sigma}}{d t'}&= (r m - \frac{v' t'}{2} \sigma) C_{m, \sigma}  \\
        & \quad - \sum_{n,\sigma'} \left [ \frac{\kappa_{mn}}{2} \delta_{\sigma, -1} \delta_{\sigma',1} + \frac{\kappa^*_{mn}}{2}  \delta_{\sigma, 1} \delta_{\sigma', -1} \right ] C_{n, \sigma'}.   \nonumber
\end{align}
We solve this system of equations numerically with a truncated oscillator basis.  First we consider the initial state of the spin system to be spin-down with the oscillator in its quantum ground state $\ket{\Psi (-\infty)} = \ket 0 \ket \downarrow$, which gives $C_{0, -1} (- \infty) = 1$ with all other $C_{m, \sigma} (- \infty) = 0$.

Strong coupling ($\lambda \sim 1$) of spin dynamics to torsional oscillations results in rich dynamics of both the spin and the oscillator.  Calculating the expectation value of $\sigma_z$,
\beq
    \langle \sigma_z \rangle = \sum_{m,\sigma} \sigma |C_{m, \sigma}|^2
\eeq
we define the probability of staying in the initial spin-down state as
\beq
    P(t) = \frac 1 2 (1 - \langle \sigma_z \rangle).
\eeq
A comparison of staying probabilities for different parameters is shown in Fig.~\ref{fig:P_ground}.  For $r \gg 1$, the spin transitions are clearly independent, as shown in Figs.~\ref{fig:P_r20_lambda1_ground} and \ref{fig:P_r20_lambda2_ground}.  The tunnel splitting at each crossing is strongly renormalized, according to Eq.~\eqref{eq:Delta_mn}, which leads to strong dependence of the transition probability on the coupling.

Consider the crossing of diabatic energies $E_{m \downarrow}$ and $E_{n \uparrow}$.  For the system initially in the $\ket m \ket \downarrow$ state, which corresponds to the lower of the two adiabatic states long before the avoided crossing, the probability that the system will stay in the initial state is
\beq
    P_{mn} = e^{-\epsilon_{mn}}, \qquad
    \epsilon_{mn} = \frac{\pi \Delta^2 | \kappa_{mn} |^2 }{2 \hbar v}.
\eeq
When the system is initially in the $\ket 0 \ket \downarrow$ state, all diabatic crossings will occur between energies $E_{0 \downarrow}$ and $E_{n \uparrow}$.  The transition probability $P_{0n} = e^{-\epsilon_{0n}}$ at each crossing depends on $|\kappa_{0n}|^2$.  Using $L_0^n(x) = 1$ we obtain
\beq
    \epsilon_{0n} = \frac{\pi \Delta^2 e^{-\lambda^2}}{2 \hbar v} \frac{\lambda^{2n}}{n!}.
\eeq
After the first avoided crossing at $t_0 = 0$, the asymptotic staying probability in the initial state is $P_{00} = e^{-\epsilon_{00}}$.  The next avoided crossing occurs at $t_1$, and the probability of staying in the spin down state after $t_1$ is $P_{01} = e^{-\epsilon_{01}}$.  Thus the total staying probability after two avoided crossings is $P_{00} P_{01}.$  We define $P_N$ as the probability of remaining in the initial state after $N$ avoided crossings,
\beq
    P_{N} = \exp{ \left ( -\sum_{n = 0}^N \epsilon_{0n} \right )}.
\eeq
In the limit $N \rightarrow \infty$, we recover the exact Landau Zener probability $P_{LZ}$,
\beq
    \lim_{N \to \infty} P_{N} = \exp{\left ( - \frac{\pi \Delta^2 e^{- \lambda^2}}{2 \hbar v} \sum_{n = 0}^\infty \frac{\lambda^{2n}}{ n!} \right ) } = \exp{\left ( - \frac{\pi \Delta^2}{2 \hbar v} \right )}.
\eeq
Fig.~\ref{fig:P_r20_lambda2_ground} shows staying probability for larger coupling, $\lambda = 2$.  We see that as the tunnel splitting at each avoided crossing is more strongly renormalized, it takes more crossings to reach the final Landau-Zener probability.

As the oscillator frequency decreases compared to the sweep rate, $r \gtrsim 1$, the transitions are no longer completely independent, although small oscillations about individual plateaus can still be seen in $P(t)$.  This is because the transitions happen within a small multiple of the  Landau-Zener tunneling time $\tau_{LZ}$.  When the oscillator frequency and tunnel splitting are close to resonance $r \sim 1$, the transition probability initially approaches $P_{LZ}$ and then shows collapse and revival behavior around this limit, as shown in Fig.~\ref{fig:P_r1_lambda1_ground}.  For $r \ll 1$ the revivals become much weaker and the probability resembles the traditional LZ probability.

\begin{figure}[!h] 
        \subfloat{\label{fig:P_CS_05_0}\includegraphics[trim = 0cm 0cm 0cm 0cm, clip, width=0.45\textwidth]{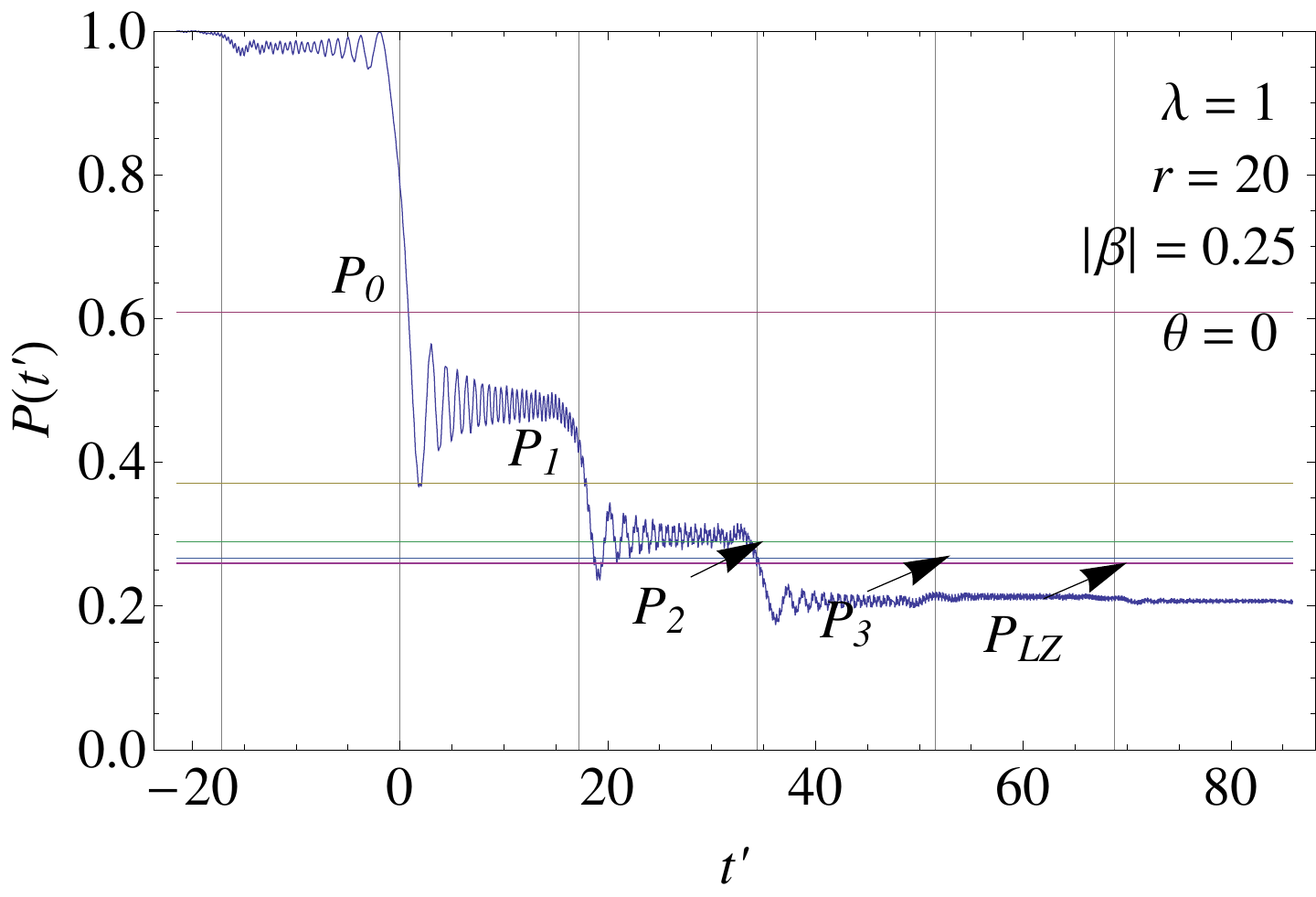}}
        \caption{Time dependence of the probability that the spin stays in the initial spin-down state for an initial coherent oscillator state $\Psi(- \infty) = \ket \beta \ket \downarrow$ with $\epsilon = 1.35$.  Vertical lines at $t_k$ denote avoided crossing of adiabatic energy levels.  Horizontal lines are exact results $P_N$ for independent transitions when starting in the $\ket 0 \ket \downarrow$ state.}
    \label{fig:P_CS}
\end{figure}

When the oscillator is initially in a coherent state $\ket \beta$,
\beq        \label{eq:coherent}
    \Psi(- \infty) = \ket \beta \ket \downarrow =
    e^{-|\beta|^2 / 2} \sum_{n = 0}^\infty \frac{\beta^n}{ \sqrt{n!} } \ket n \ket \downarrow
\eeq
where the complex number $\beta = |\beta| e^{i \theta}$ is proportional to the amplitude of initial oscillations.  When $\beta \ll 1$ the spin transitions follow approximately the same asymptotic values $P_N$ as the quantum ground state case.  For $\beta \lesssim 1$ the staying probabilities, an example of which is shown in Fig.~\ref{fig:P_CS}, depend on the magnitude and phase of the initial coherent state.  The maximum angular displacement and velocity of a coherent state are related to $\beta$ through
\beq		\label{eq:phi_beta}
	\varphi_{\mathrm{max}} = 2 \lambda | \beta |, \qquad
	\left ( \frac{d \varphi}{dt'} \right )_{\mathrm{max}} = 2 r \lambda | \beta |,
\eeq
where $\varphi = 2 S \langle \phi \rangle$.

%
\subsection{Weak coupling \label{subsec:weak}}

When the spin dynamics of the nanomagnet are weakly coupled to its rotational dynamics $\lambda \ll 1$, there is little observable effect of rotations on spin flip probability.  The first crossing that occurs at $t_0 = 0$ has tunnel splitting $\Delta_{00} = \Delta e^{-\lambda^2/2}$, which tends to unity for small $\lambda$.  When $r \gg 1$ the first transition at $t_0 = 0$ approaches $P_{00} = e^{- \epsilon_{00}}$.  The second independent transition occurs at $t_1$ approaches $P_{LZ}$, although the difference between $P_{00}$ and $P_{LZ}$ is very small.  When $r \sim 1$ the adiabatic transitions are no longer independent, and occur close to the Landau-Zener tunneling time interval.

%
%
\section{Oscillator dynamics \label{sec:osc}}
We compute the expectation value of the torsional rotation angle as a function of time
\beq
   \varphi = \lambda \sum_{m, \sigma} \left ( C_{m + 1, \sigma}^* C_{m, \sigma} \sqrt{m + 1} + C_{m - 1, \sigma}^* C_{m, \sigma} \sqrt{m} \right ).
\eeq
\begin{figure}[h!] 
        \subfloat{\label{fig:phi_r20_lambda1_ground}\includegraphics[trim = 0cm 0cm 0cm 0cm, clip, width=0.43\textwidth]{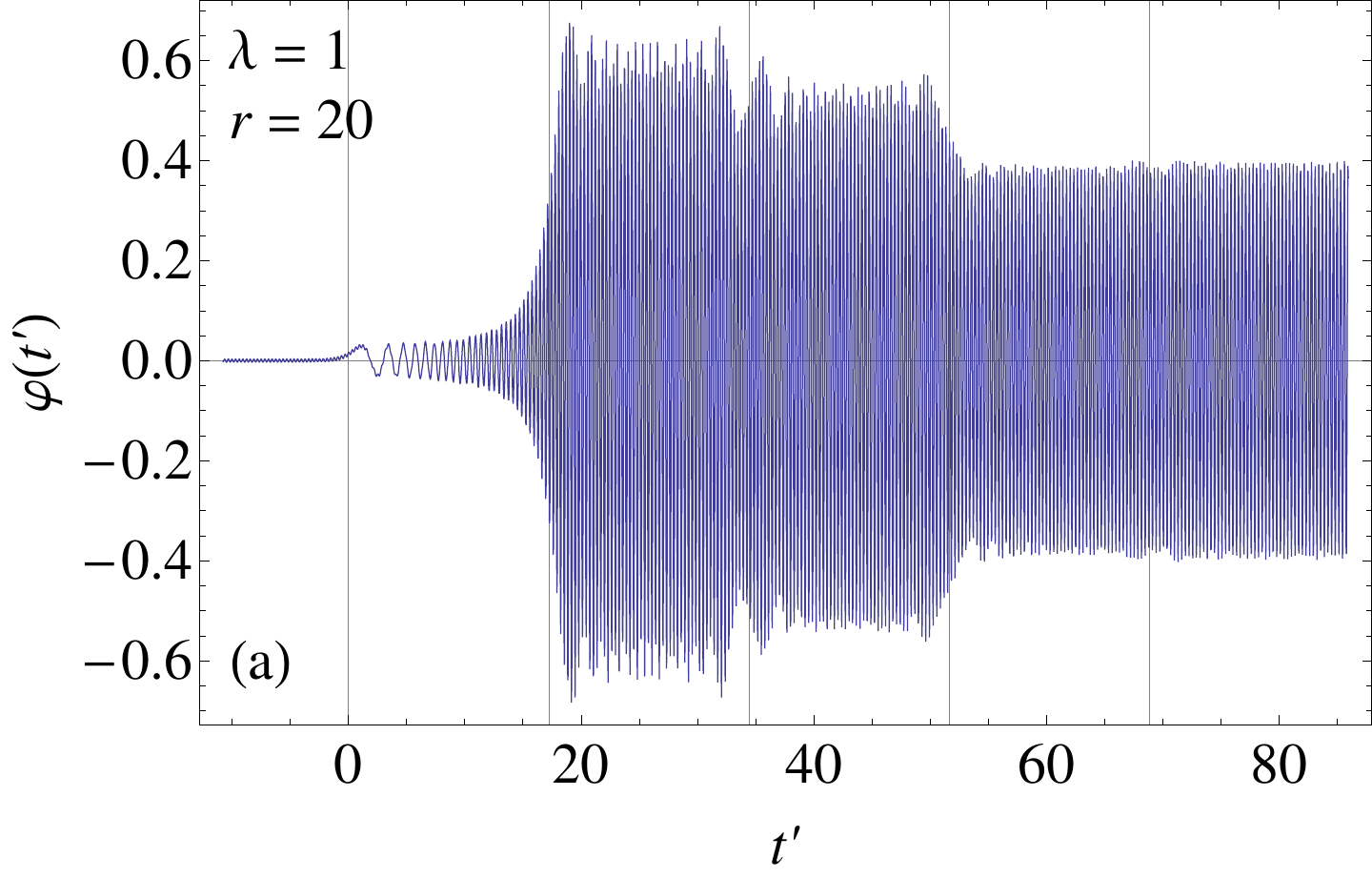}}  \\
        \subfloat{\label{fig:phi_r20_lambda2_ground}\includegraphics[trim = 0cm 0cm 0cm 0cm, clip, width=0.43\textwidth]{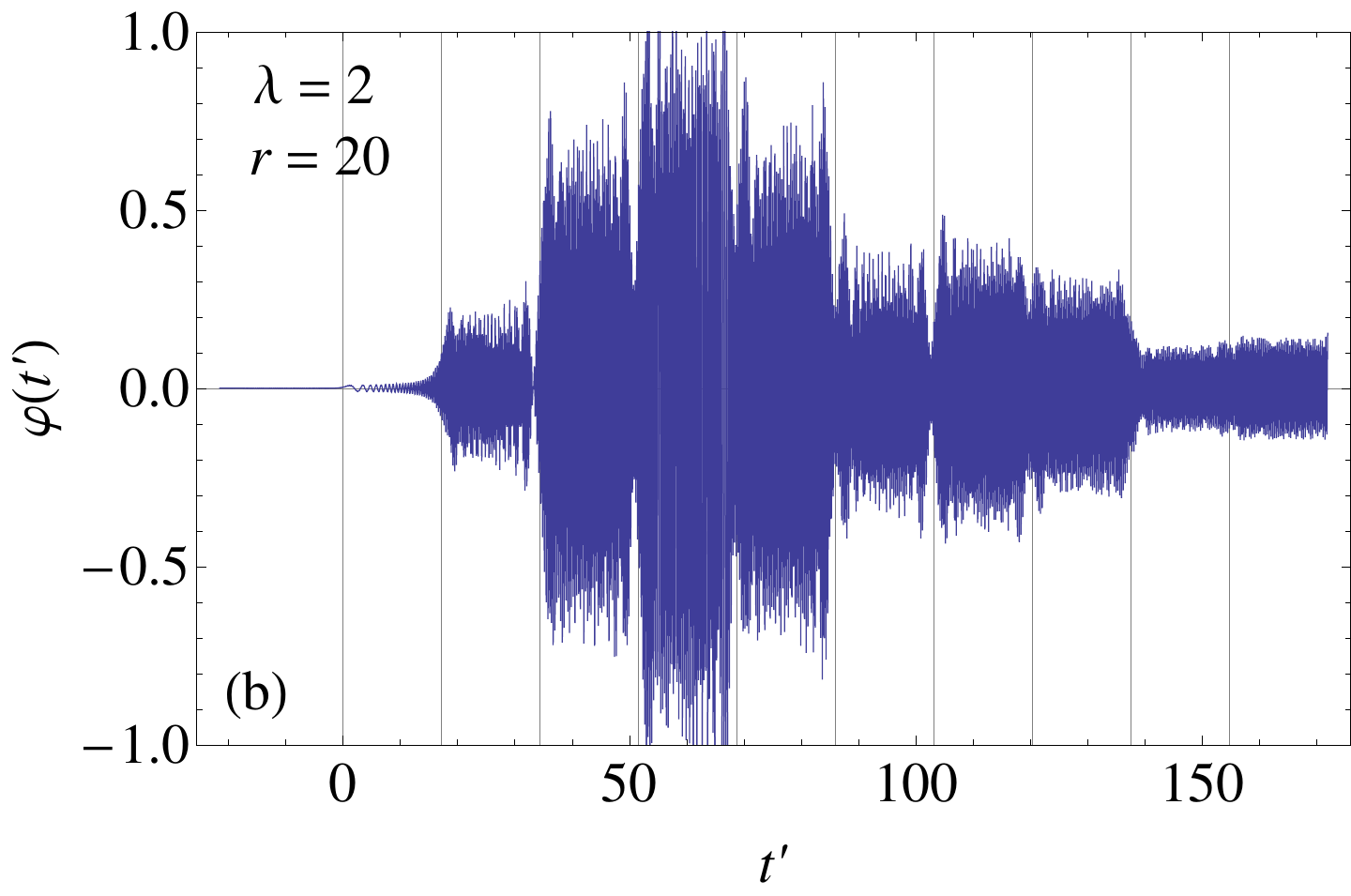}}  \\
        \subfloat{\label{fig:phi_r1_lambda1_ground}\includegraphics[trim = 0cm 0cm 0cm 0cm, clip, width=0.43\textwidth]{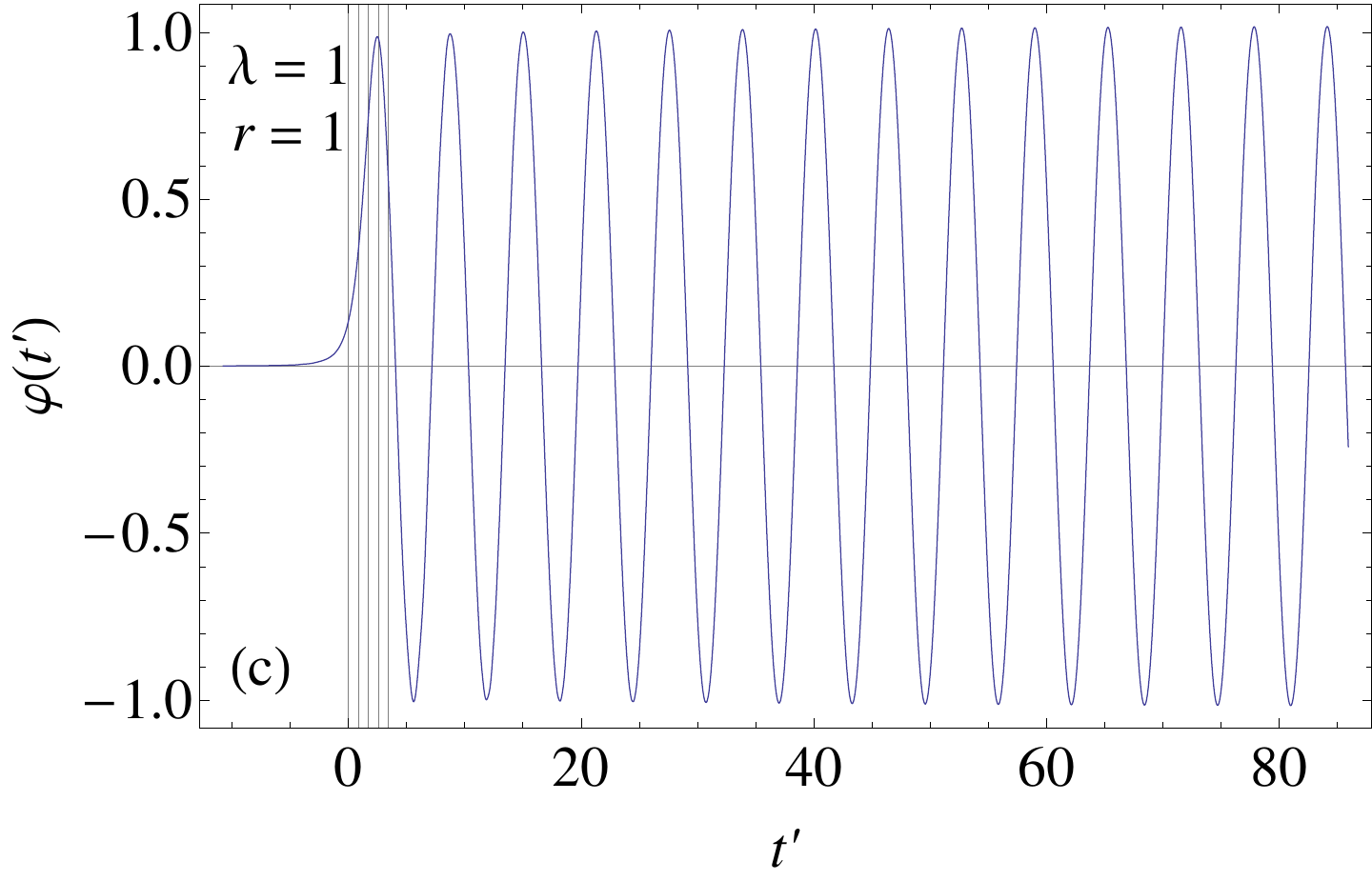}}
        \caption{Time dependence of the rotation angle expectation value for initial state $\Psi(- \infty) = \ket 0 \ket \downarrow$ with $\epsilon = 1.35$.  Vertical lines at $t_k$ denote avoided crossing of adiabatic energy levels.  }
    \label{fig:phi_ground}
\end{figure}
\begin{figure}[h!]  
        \subfloat{\label{fig:phi_CS_05_0}\includegraphics[trim = 0cm 0cm 0cm 0cm, clip, width=0.45\textwidth]{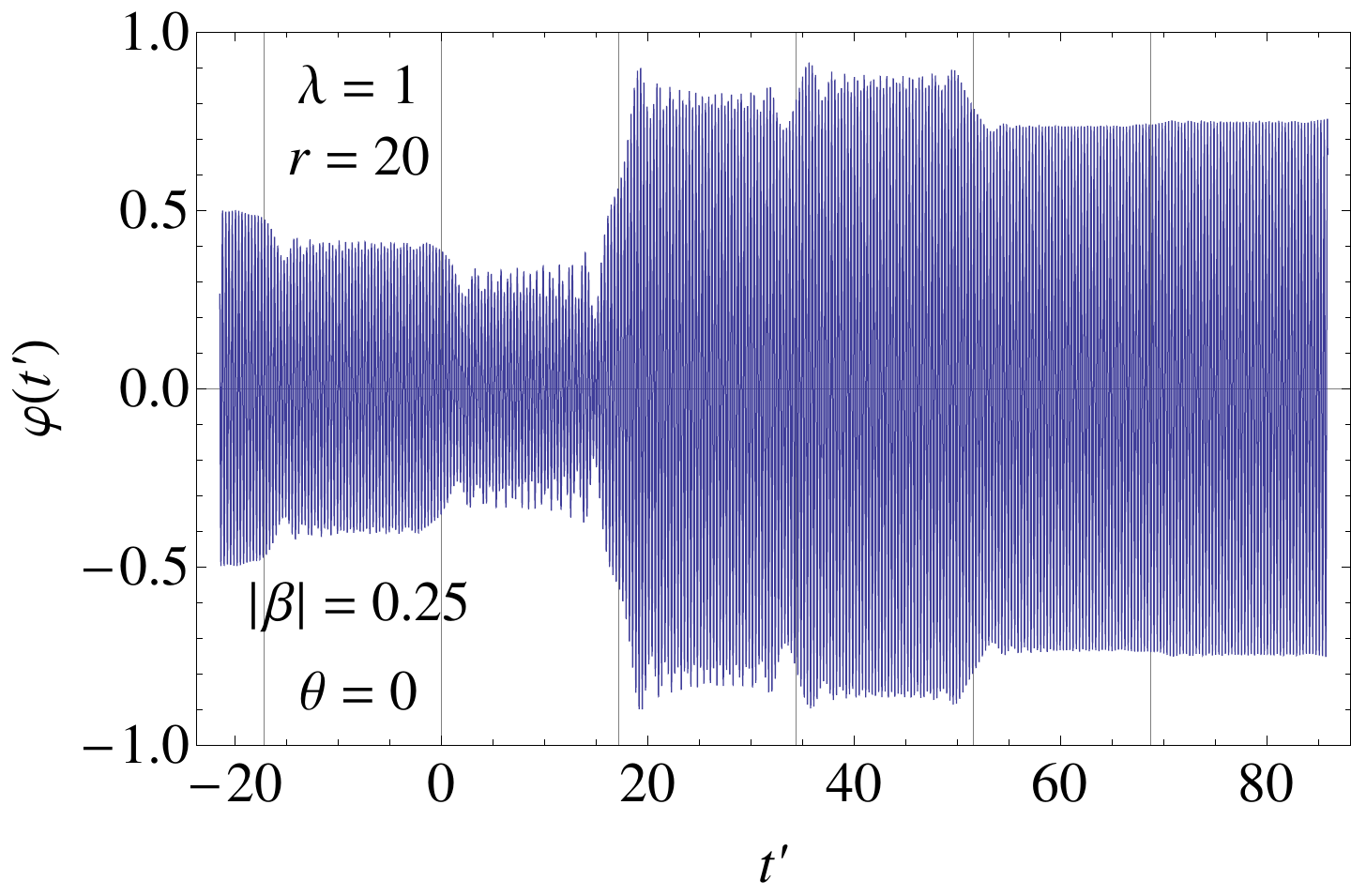}}
        \caption{Time dependence of the rotation angle expectation value for initial coherent state $\Psi(- \infty) = \ket \beta \ket \downarrow$ with $\epsilon = 1.35$.  Vertical lines at $t_k$ denote avoided crossing of adiabatic energy levels.  }
        \label{fig:phi_CS}
\end{figure}
For strong coupling $\lambda \sim 1$, the dynamics of the resonator shows a delay before the onset of large oscillations for $r \gg 1$, which occurs at $t_1$, shown in Fig.~\ref{fig:phi_r20_lambda1_ground}.  When the coupling is stronger, Fig.~\ref{fig:phi_r20_lambda2_ground} shows many more changes in the oscillatory motion, consistent with more avoided crossings.  The oscillation amplitude changes slightly at subsequent $t_k$.  As $r$ decreases towards $1$, the interval of large oscillations becomes shorter.  When $r \lesssim 1$, there is a single transition region which gives way to harmonic oscillations, shown in Fig.~\ref{fig:phi_r1_lambda1_ground}.  Near $r = 1$, the amplitude of oscillations tends to increase as $r$ decreases for fixed $\lambda$.

The angular displacement of the torsional resonator also shows interesting effects even for small coupling.  When $r \gg 1$, large torsional oscillations do not begin at the first crossing.  This can be understood as follows.  The $t = 0$ crossing occurs between spin up and down states, both of which correspond to the ground state of the resonator.  Although there is a small increase in displacement angle at this crossing, the largest increase occurs at the second crossing between $\ket 0 \ket \downarrow$ and $\ket 1 \ket \uparrow$ at $t_1$.  This delay agrees exactly with the semiclassical treatment by Jaafar et al. \cite{Jaafar:2010}.  Following this time the oscillator is in a superposition of ground and excited states.  When $r \lesssim 1$ successive transitions occur in a short duration compared to $\tau_{LZ}$, and there is no observable delay in the onset of oscillations.  The oscillation amplitude depends on the sweep rate.  Numerical results suggest that the largest amplitude oscillation reaches a maximum near $\epsilon \simeq 2$ for $\lambda = 1$ and $r = 20$.

When the oscillator is initially in a coherent state $\ket \beta$, normal oscillations with maximum amplitude $\varphi_{\mathrm{max}} = 2 \lambda |\beta|$ occur up to $t_{-1}$, as shown in Fig.~\ref{fig:phi_CS}.  At $t_{-1}$ the amplitude decreases slightly and decreases again at $t_0$.  A large increase occurs at $t_{1}$, similar to the case where the oscillator is initially in its quantum ground state.  The amplitude of oscillations after $t_1$ tends to be larger when the oscillator is initially in a coherent state, but not by a large amount.  We observe a subsequent change in oscillation amplitude at $t_2$ and $t_3$.  The oscillator dynamics are not as sensitive to the initial phase of the coherent state as the spin dynamics, although there is some variation in maximum amplitude.

%
%
\section{Collective dynamics of spins coupled to a mechanical resonator \label{sec:SR}}
Consider, instead of a single nanomagnet, an array of single molecule magnets with their easy axes mutually aligned with the axis of rotation of the resonator.  If they are far enough apart that dipolar coupling is negligible, they will only be coupled through the effective field due to torsional oscillations.  Because the angular displacement is the same for each molecule, this results in collective coherent dynamics, described by a variant of the Dicke Hamiltonian.  For $N$ single molecule magnets, we define the operator of total low-energy dynamics as
\beq
    \hat H_R = - \frac \Delta 2 R_x, \qquad
    \mathbf R = \sum_{i=1}^N \boldsymbol \sigma^{i}
\eeq
where the index $i$ labels each magnetic particle.  Again transforming to the lab frame by performing a rotation by angle $\phi$ to the lab frame, but now using the total spin, we obtain
\begin{align}
    \hat H_R'   &= - \frac \Delta 2 \left ( e^{- 2 i S \phi} R_+ + e^{2 i S \phi} R_-  \right )  \nonumber  \\
    &= - \frac \Delta 2 \left ( \cos{(2 S \phi)} \, R_x + \sin{(2 S \phi)} \, R_y \right ).
\end{align}
The full Hamiltonian for the array of single molecule magnets is
\begin{align}
    \hat H_{SR} &= \frac{\hbar^2 L_z^2}{2 I_z} + \frac{I_z \omega_r^2 \phi^2}{2} - \frac{W (t)}{2} R_z  \nonumber  \\
        & \qquad  - \frac \Delta 2 \left ( \cos{(2 S \phi)} \, R_x + \sin{(2 S \phi)} \, R_y \right ),
\end{align}
where $W(t) = v t$.  The Hamiltonian can be written as
\beq
    \hat H_{SR} = \hat H_{\mathrm{osc}} - \frac 1 2 \mathbf H_{\mathrm{eff}} \cdot \mathbf R,
\eeq
where $\hat H_{\mathrm{osc}}$ is the uncoupled oscillator Hamiltonian and
\beq
    \mathbf H_{\mathrm{eff}} = - \frac{\delta \hat H}{\delta \mathbf R} = \Delta \cos{(2 S \phi)}\mathbf e_x + \Delta \sin{(2 S \phi)} \mathbf e_y + W  \mathbf e_z
\eeq
is the effective magnetic field.  Noticing that $\hat H_{SR}$ is linear in $R_x, R_y, R_z$, we can see that $[\mathbf R^2, \hat H_{SR} ] = 0$, so $\mathbf R^2 = R (R+1)$ is a conserved quantum number and $\mathbf R$ behaves as a single large isospin.  We are interested in the maximum value of $R$, $R_{\mathrm{max}} = N / 2$, which can be experimentally realized by preparing the system with a strong longitudinal magnetic field such that all spins are pointing down.

The Heisenberg equations of motion $i \hbar \, d \hat A / d t = [\hat A, \hat H] $ are
\begin{align}
    \hbar \dot L_z &= - I_z \omega^2 \phi - \Delta S \left ( \sin{(2 S \phi)} R_x - \cos{(2 S \phi)} R_y \right )   \label{eq:Ldot} \\
    \dot \phi &= \frac{\hbar L_z}{I_z}  \label{eq:phidot}  \\
    \hbar \dot R_x &= W R_y - \Delta \sin{(2 S \phi)} R_z  \label{eq:Rxdot}  \\
    \hbar \dot R_y &= - W R_x + \Delta \cos{(2 S \phi)} R_z  \label{eq:Rydot}  \\
    \hbar \dot R_z &= - \Delta \cos{(2 S \phi)} R_y + \Delta \sin{(2 S \phi)} R_x.  \label{eq:Rzdot}
\end{align}
The equations of motion show a few important properties.  First, the time derivative of the $z$-component of the total angular momentum equals the elastic torque
\beq
    \frac{d}{dt} (\hbar L_z + \hbar S R_z) = - I_z \omega^2 \phi.
\eeq
If the spin-rotor system were completely uncoupled from its environment the total angular momentum, spin plus rotational, would be conserved.  In the limit $\phi \rightarrow 0$, we would obtain Heisenberg equations of motion for $R_{x,y,z}$.  Solving this system of equations gives the same Landau-Zener probability of spin flip as the Schr\"odinger picture, discussed in Sec.~\ref{subsec:LZ}.
Second, these equations are not independent, but
\beq
    \frac{d}{dt} \mathbf R^2 = 0
\eeq
which is equivalent to $ \mathbf R^2 = \text{constant}$, which we had found as a constant of motion of the Hamiltonian.  Because the length of $\mathbf R$ is fixed and large in magnitude, we see that the equations of motion for $\mathbf R$ are equivalent to the Landau-Lifshitz equations for a classical spin of fixed length precessing in a magnetic field.  Dividing Eqs.\eqref{eq:Ldot}-\eqref{eq:Rzdot} by $R$ shows that the direction of the total spin follows the Landau-Lifshitz equation, which is mathematically equivalent to the Schr\"odinger equation of a spin-half particle precessing in a magnetic field,
\beq
    \hbar \frac{d \boldsymbol \sigma}{d t} = \boldsymbol \sigma \times \mathbf H_{\mathrm{eff}}, \qquad \boldsymbol \sigma = \frac{\mathbf R}{R}.
\eeq
The equations of motion for $R_{x,y,z}$ can be divided through by R to giving identical equations of motion for a pseudospin $\boldsymbol \sigma = \mathbf R / R$ of unit length.  Substituting this into the equation of motion for $\phi$ and eliminating $L_z$ gives a second order equation of motion for for the dynamics of the resonator,
\beq        \label{eq:phi_de}
    \frac{d^2 \varphi}{d t'^2} + r^2 \varphi = - \alpha R \left ( \sin{(\varphi)} \sigma_x - \cos{(\varphi)} \sigma_y \right ),
\eeq
where $\varphi = 2 S \phi$, the prime denotes derivative with respect to dimensionless time $t' = \Delta t / \hbar$, $r$ and $\alpha$ are defined in Eqs. \eqref{eq:lambda_r} and \eqref{eq:alpha}, respectively.

The right hand side of Eq.~\eqref{eq:phi_de} shows that the spins
exert a collective torque on the resonator.  This is a simple yet
meaningful result. The equation of motion is similar to the
semiclassical treatment of a single spin \cite{Jaafar:2011}, but
with the torque on the resonator increased by a factor of $R$.
Because the amplitude of oscillation is proportional to the number
of magnetic molecules $N$, this can be interpreted as a signature
of Dicke phonon superradiance \cite{Chudnovsky:2004}. For a simple
harmonic torsional oscillator, the phonon field is the angle of
displacement from equilibrium $\phi$, and the driving torque is
proportional to $R = N/2$.

Returning to the quantum model we see that for the case of
superradiance, $\alpha \rightarrow \alpha R$, and $\lambda =
\sqrt{\alpha / r}$ becomes \beq
    \lambda_{SR} = \sqrt{\frac{\alpha R}{r}} = \sqrt R \, \lambda \propto \sqrt N \, \lambda.
\eeq This provides a viable method of increasing the coupling in a
realistic experiment, by increasing the number of individual
nanomagnets on the resonator.  The usual difficulty of realizing
strong coupling is that reducing the moment of inertia by even two
orders of magnitude has a small effect on the coupling due to the
inverse quartic root dependence of the coupling on the moment of
inertia.

The Heisenberg equations of motion,
Eqs.~\eqref{eq:Ldot}-\eqref{eq:Rzdot} are operator equations which
should be averaged over the quantum state of the system.  Since the spin $\bf R$ is classical the averages decouple, such as $\langle \sin(\varphi) \sigma_x \rangle
\rightarrow \langle \sin(\varphi) \rangle \langle \sigma_x
\rangle$ in Eq.~\eqref{eq:phi_de}, which yields classical-like equations of motion.  We solve these equations of motion numerically.
\begin{figure}[h!]    
        \subfloat{\label{fig:P_r20_lambda1_SR_0}\includegraphics[trim = 0cm 0cm 0cm 0cm, clip, width=0.45\textwidth]{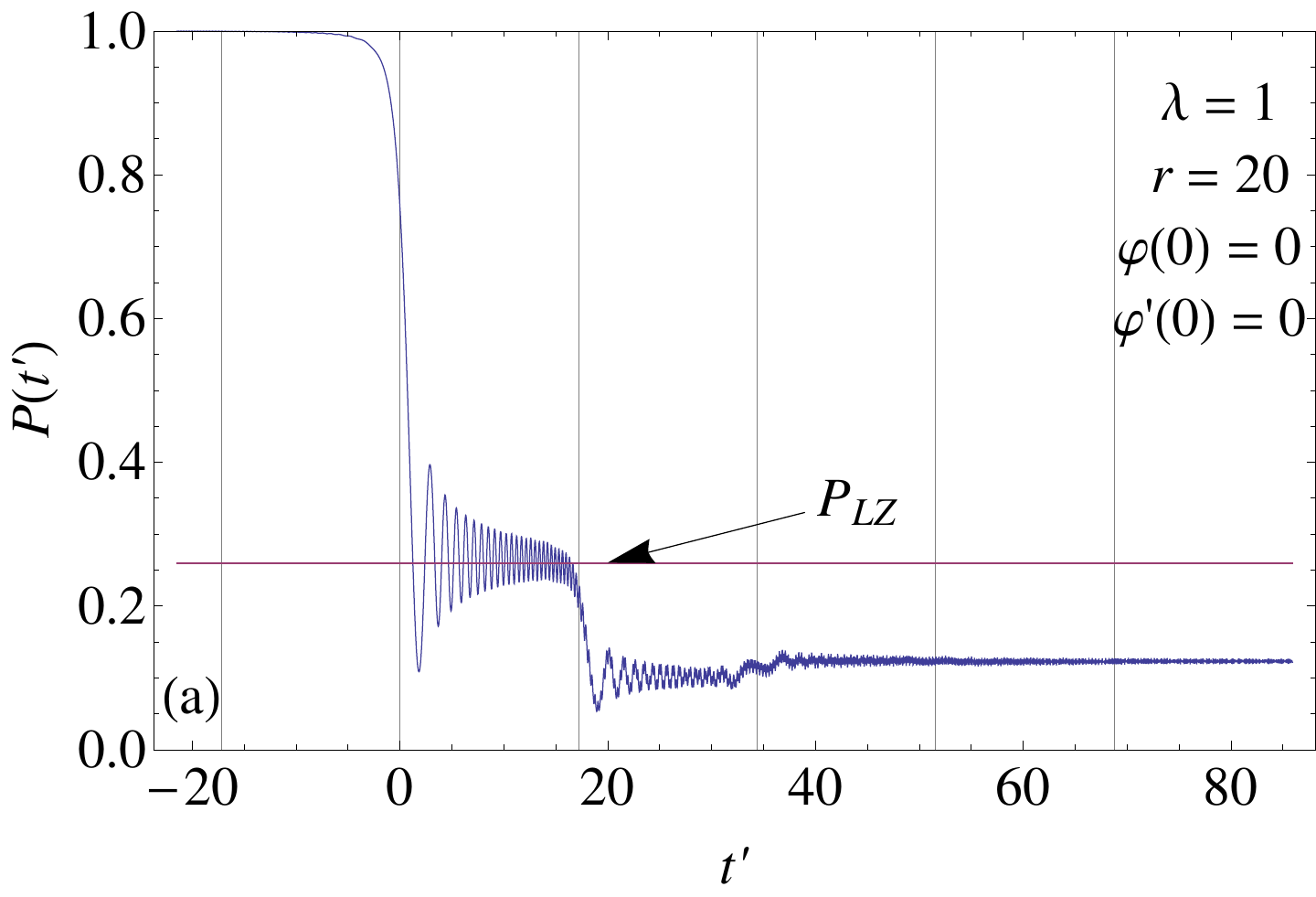}}  \\
        \subfloat{\label{fig:P_r20_lambda1_SR_05_0}\includegraphics[trim = 0cm 0cm 0cm 0cm, clip, width=0.45\textwidth]{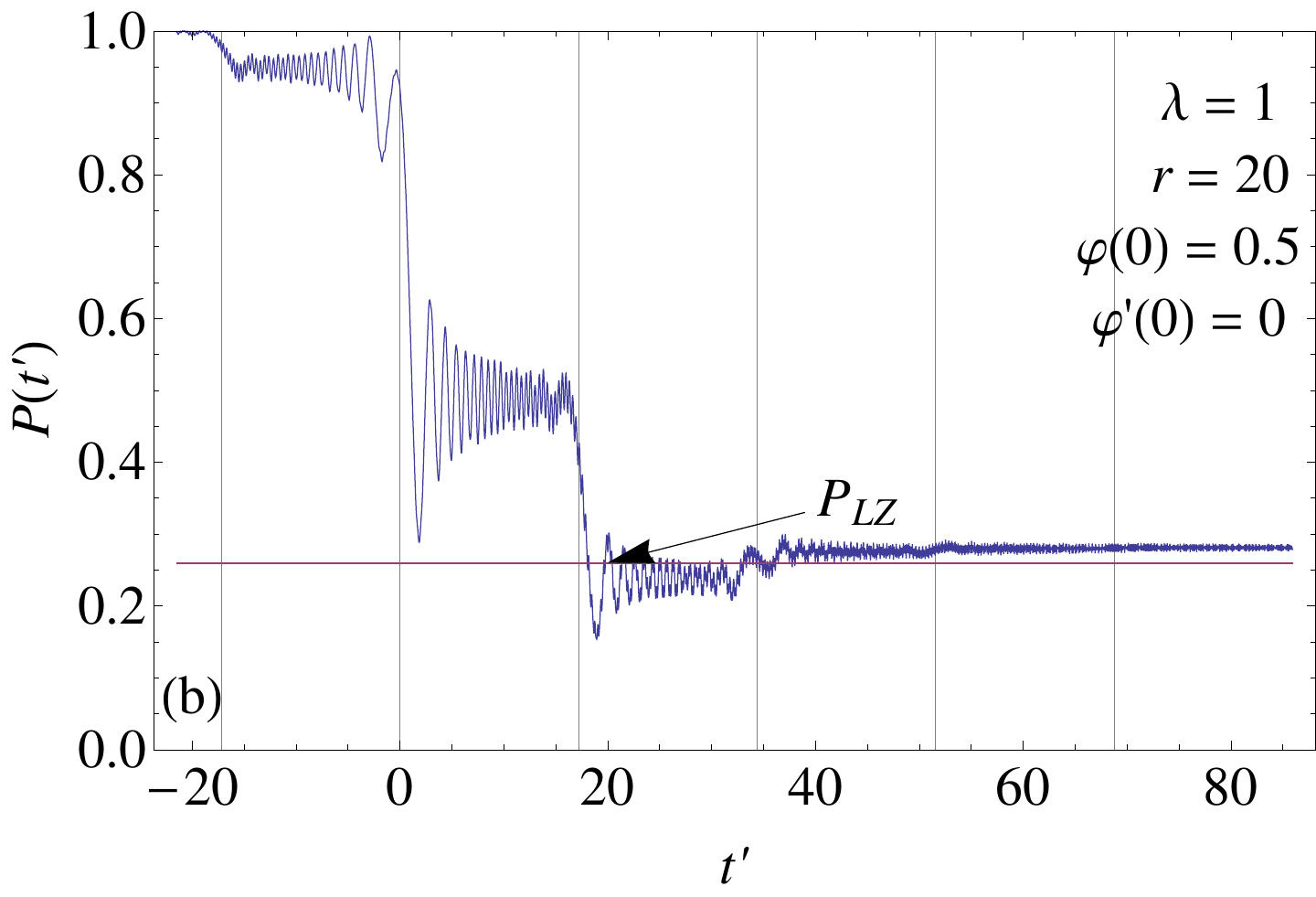}}
        \caption{Time dependence of the effective probability $P(t) = (1 - \langle \sigma_z \rangle) / 2$ for the $z$-component of a large spin in the semiclassical model, with various initial conditions and $\epsilon = 1.35$.  }
    \label{fig:P_SR}
\end{figure}

We emphasize that a large spin will display the classical dynamics of a magnetic moment precessing in a time-dependent magnetic field.  Plots of the probability as a function of time for the semiclassical equations of motion of a superradiant ensemble of spins are shown in Fig.~\ref{fig:P_SR}.  We see multi-stage transitions similar to the quantum case.  A semiclassical explanation is as follows.  Transitions occur when the energy separation between spin states equals a multiple of the oscillator frequency.  These occur at the same times $t_k = r / v'$ given by Eq.~\eqref{eq:t_k} for the quantum case when avoided crossings between adiabatic energies occur.

When the oscillator is initially at rest at its equilibrium position $\varphi = 0$, the initial transition occurs at $t_0$ as shown in Fig.~\ref{fig:P_r20_lambda1_SR_0}.  $P(t)$ oscillates about the regular Landau-Zener probability $P_{LZ} = e^{- \epsilon}$.  Subsequent transitions occur at $t_1$ and $t_2$, with the long-time probability much different from $P_{LZ}$.  The spin dynamics depend strongly on the initial state of the oscillator.  Fig.~\ref{fig:P_r20_lambda1_SR_05_0} shows the transition probability for different initial conditions of the oscillator with the same amplitude of oscillation as the coherent state studied in the fully quantum-mechanical model.

\begin{figure}[ht!]    
        \subfloat{\label{fig:phi_r20_lambda1_SR_0}\includegraphics[trim = 0cm 0cm 0cm 0cm, clip, width=0.45\textwidth]{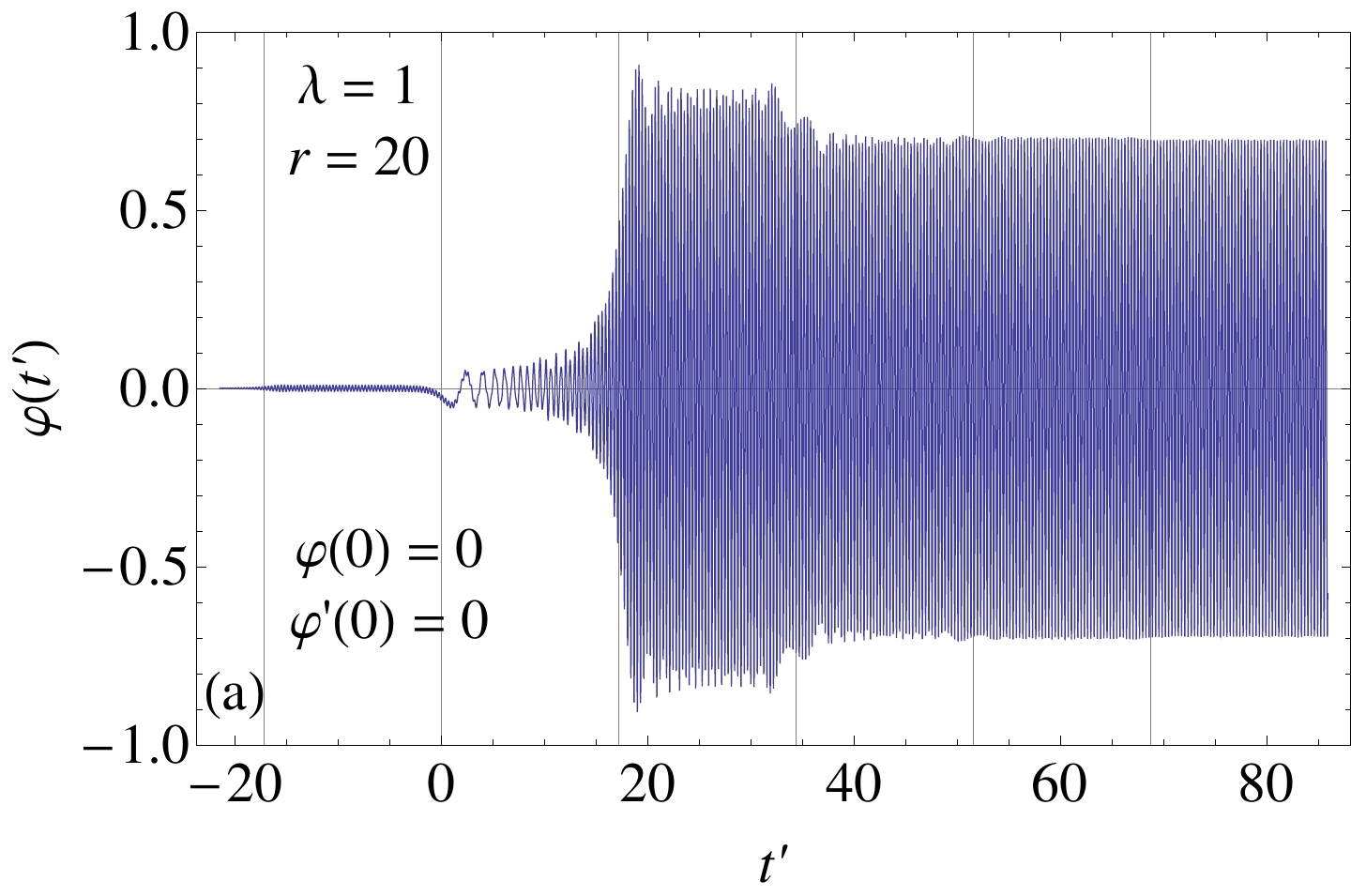}}  \\
        \subfloat{\label{fig:phi_r20_lambda1_SR_05_0}\includegraphics[trim = 0cm 0cm 0cm 0cm, clip, width=0.45\textwidth]{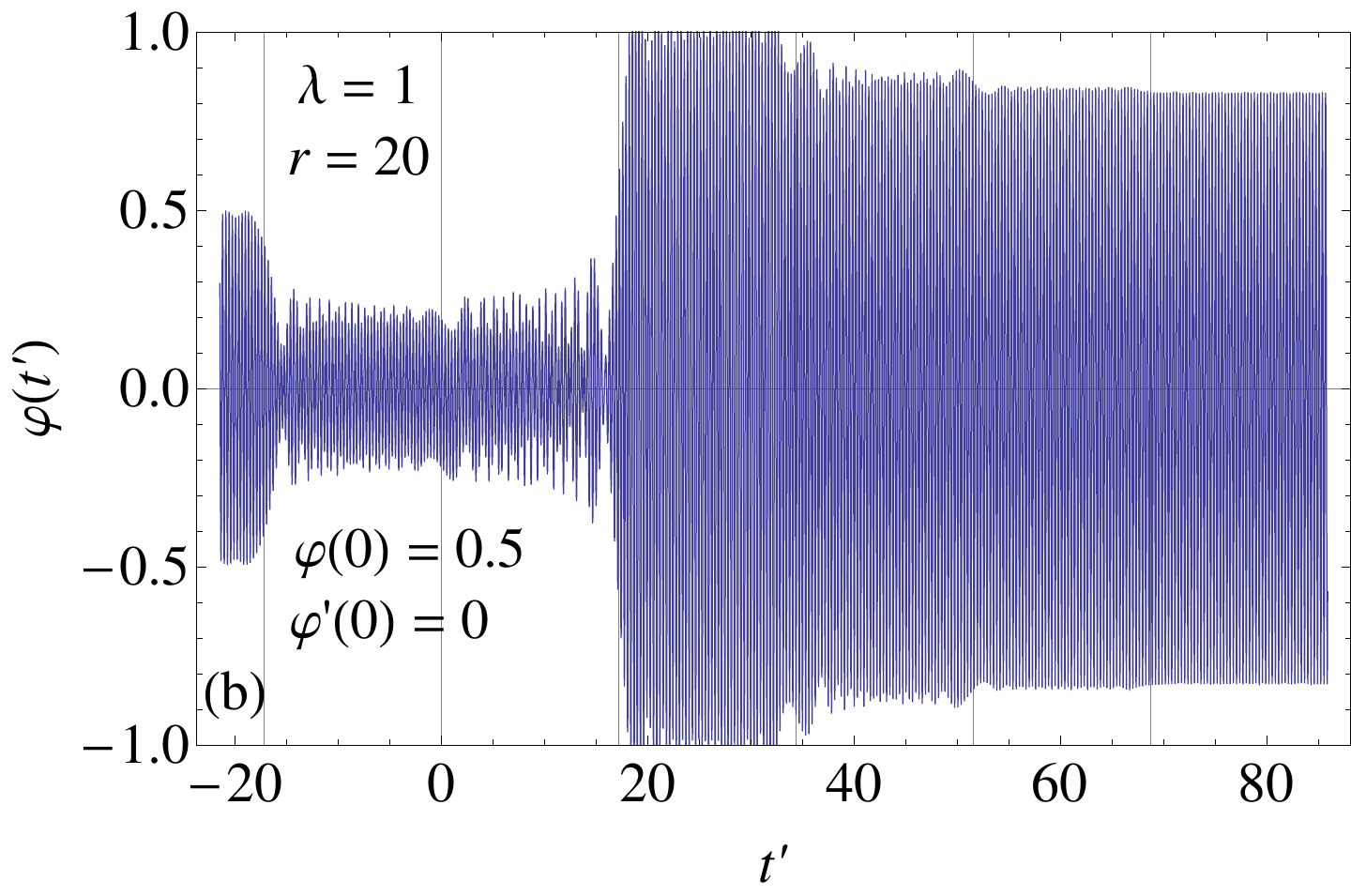}}
                \caption{Time dependence of the rotation angle in the semiclassical model, with various initial conditions and $\epsilon = 1.35$.  }
    \label{fig:phi_SR}
\end{figure}

The oscillator dynamics show a similar delay in the onset of strong oscillations as in the quantum model, shown in Fig.~\ref{fig:phi_SR}.  We notice a large increase in the amplitude of oscillations at $t_1$ for the cantilever initially at rest at equilibrium.  For the cantilever initially oscillating with amplitude $\varphi_0$, the behavior is very similar to the quantum model with the oscillator initially in a coherent state.  We observe normal oscillations up to $t_{-1}$ at which the amplitude decreases, then a large increase at $t_1$, with subequent changes at $t_2$, $t_3$.   While there is not a strong dependence on the initial conditions with the same initial energy, there is some variation in maximum amplitude.

%
%
\section{Discussion and conclusions \label{sec:conclusion}}

We have studied the Landau-Zener dynamics of a tunneling spin
rigidly coupled to a torsional oscillator.  Starting with a
quantum model describing the low energy dynamics of a tunneling
macrospin, we numerically solve the time-dependent Schr\"odinger
equation to obtain the dynamics of the expectation values of the
spin and oscillator.  We find that when the oscillator is
initially in its quantum ground state, there are a series of
plateaus in the staying probability as a function of time.  We
analytically obtain exact probabilities in terms of tunnel
splittings of the spin which are dressed by the quantum states of
the torsional oscillator.  These results perfectly fit the
plateaus obtained from numerical simulations.  The oscillator
dynamics show abrupt changes in amplitude which occur at the same
times as the steps between steps of the staying probability.  For
an oscillator initially in a coherent state we also find a
stepwise staying probability curve, but these deviate from the
analytical results found for the initial ground state because
there are multiple occupied states of the resonator.  The
oscillator dynamics continue to show changes in amplitude which
coincide with the steps.  We also consider a large number of
spins, $N$, on a single oscillator, and find a superradiant
enhancement of the spin-oscillator coupling which scales as $\sqrt
N$. As in the Dicke model, the ensemble of spins acts as a single
large spin. This justifies decoupling quantum averages of separate
observables in the Heisenberg equations of motion, giving
semiclassical equations of motion for a large spin in a
time-dependent effective field which depends on the motion of the
cantilever.  The cantilever experiences a harmonic restoring
torque but also a driving torque due to the dynamics of the large
spin.  We numerically solve the set of coupled equations and
compare the results to the Schr\"odinger picture.  The spin
dynamics show sensitivity to the initial state of the resonator,
although the oscillator dynamics are fairly insensitive to this.

It is important to distinguish the interpretation of these results
in the context of the system being measured.  Consider the single
molecule magnet grafted to a carbon nanotube, depicted in Fig.~\ref{fig:CNT_SMM}.  With the system prepared in the spin down
state by a strong magnetic field along the negative $z$-axis, the
magnetic field is swept.  If the oscillator is initially in the
zero phonon state, the first crossing of an occupied energy level
with an unoccupied level occurs at $t_0 = 0$ between $E_{0
\downarrow}$ and $E_{0 \uparrow}$.  $P_{0} = P_{00} =
e^{-\epsilon_{00}}$ is the probability that the spin will remain
in the down state.  If the spin remains in the down state after
the first crossing, it will encounter a second crossing between
$E_{0 \downarrow}$ and $E_{1 \uparrow}$ at $t_1$, at which it will
remain spin-down with probability $P_{01} = e^{-\epsilon_{01}}$.
The total probability of the spin remaining spin-down after $t_1$
is $P_1 = P_{00} P_{01} = e^{- (\epsilon_{00} + \epsilon_{01})}$.
If the spin reverses at any $t_k$ it will see no more crossings.
When the spin does reverse it will exert a torque on the carbon
nanotube, exciting a phonon mode.  The onset of oscillations shows
that the spin has tunnelled. This provides a method of detecting
the mechanical quantum state of the nanotube.

This situation is similar to the recent demonstration of
electronic readout of nuclear spin states of a terbium-based
single magnetic molecule \cite{Vincent:2012}. Terbium nuclear spin
$3/2$ has four possible projections onto the quantization axis,
each projection providing a different hyperfine shift of the
resonance of the Landau-Zener transition of the spin of the
molecule. Time-resolved measurements show an increase in the
differential conductance at the time the spin makes a transition.
This occurs at a different value of the external field for each
sweep, that depends on the quantum state of the nuclear spin. In
our model the role of nuclear spin states is played by the
resonator states given by Eq.~\eqref{eq:E_n_updown}. We,
therefore, propose a similar experiment in which the field sweep
is used to read out the quantum state of the mechanical resonator.

When there is a large number of magnetic molecules on a
cantilever, as in Fig.~\ref{fig:cantilever_SMMs}, they will act as a single large classical spin ${\bf
R}(t)$. This spin not only responds to the external field but also
to the motion of the cantilever. The latter has been treated in
Sec. V as a classical oscillator described by the angle $\phi(t)$.
Such treatment is the classical limit of the quantum-mechanical
consideration in which the cantilever is described by the coherent
state, Eq.~\eqref{eq:coherent}. Mechanical rotation at an angular
frequency $\dot \phi$ is equivalent to a magnetic field
$B_{\mathrm{eff}} = \dot \phi / \gamma$, where $\gamma$ is the
gyromagnetic ratio.  In turn, the spin dynamics act as a driving
torque on the cantilever, resulting in coupled dynamics which
change at the same moments of time, $t_k$, as in the quantum case.
The non-linear coupled equations of motion lead to the excitations
of harmonics of the cantilever that correspond to its quantum
modes in the dynamics described by the Schrodinger equation.
Higher harmonics are excited with smaller amplitude.

To put some of these statements into perspective, consider a
spin-10 single molecule magnet grafted to a carbon nanotube
\cite{Ganzhorn:2013}.  The moment of inertia of the magnetic
molecule is of the order $I_z \sim 10^{-42}$ kg$\cdot$m$^2$.  With a
carbon nanotube torsional stiffness of $k \sim 10^{-18}$ N$\cdot$m
the simple harmonic model gives $\omega_r \sim 1000$ GHz, which
means coupling on the order of $\lambda \sim 10^{-1}$.  Typical
phonon frequencies of carbon nanotubes in the 10-100 GHz range
would increase the coupling by an order of magnitude.  Recent
observation \cite{Ganzhorn:2013} of strong spin-phonon coupling in
such a system estimates $\lambda \simeq 0.5$.  While this is
certainly large enough to observe the influence of the oscillator
on spin dynamics, there is no way to directly observe oscillations
in a carbon nanotube.

If the same spin-10 magnetic molecule were mounted on a
paddle-shaped torsional resonator of size $20 \times 20 \times 10$
nm$^3$ supported by a single carbon nanotube with torsional
rigidity $k = 10^{-18}$ N$\cdot$m.  The moment of inertia is
dominated by the paddle, $I_z \sim 10^{-36}$ kg$\cdot$m$^2$, which
gives $\omega_r = \sqrt{k / I_z} \sim 10^9$ s$^{-1}$.  The
coupling parameter $\lambda$ is then on the order of $10^{-2}$,
which would be too small to observe an effect on the spin
dynamics.  With $\Delta / \hbar \ll 10^{9}$ s$^{-1}$ there should
be a detectable delay between the $t = 0$ crossing and the onset
of maximal oscillation amplitude.  With $\Delta / \hbar > 10^{9}$
s$^{-1}$, the delay will be undetectable.  The tunnel splitting
can be tuned by orders of magnitude by applying a transverse
magnetic field.

A macroscopic resonator in which even small amplitude oscillations
could be observed comes at the expense of weak coupling with no
observable effect on the spin dynamics.  In terms of the moment of
inertia and torsional stiffness, the coupling goes as $\lambda
\propto 1 / \sqrt[4]{k I_z}$, so a very small torsional stiffness
of $k \sim 10^{-22}$ N$\cdot$m would be needed.  One way to
overcome this limitation is to put a large number of spins on a
torsional resonator or microcantilever. For a cantilever with
dimensions $1000 \times 200 \times 100$ nm$^3$ we would expect
$\omega \sim 1$ GHz with $Q \sim 500$. Single molecule magnets
have a diameter on the order of 1 nm.  It would be possible to
place hundreds of single molecule magnets on the tip of a
nanocantilever separated by over 10 nm from their nearest
neighbors to weaken dipolar interactions.  They would act as a
single large spin due to the collective quantum effect of
superradiance. This would increase the coupling by at least an
order of magnitude, as $\lambda_{SR} \propto \sqrt N$.  Therefore
it would be possible to directly observe the coupled dynamics of
the magnetization and oscillatory motion in a Landau-Zener
experiment.

\begin{acknowledgments}
This work has been supported by the U.S. National Science
Foundation through grant No. DMR-1161571.
\end{acknowledgments}

%
\bibliography{LZ_biblio}

\end{document}